\documentclass[final,12pt]{clear2025} 

\title[Causal-Effect Score]{The Causal-Effect Score in Data Management\\
{\small (slightly extended and revised version of the CLeaR'25 paper in JMLR  275:874–893, 2025)}}
\usepackage{times}

\clearauthor{%
 \Name{Felipe Az\'ua} \Email{fazua@alumnos.uai.cl}\\
 \addr Universidad Adolfo Ib\'a\~nez, Santiago, Chile.%
 \AND
 \Name{Leopoldo Bertossi}\thanks{Professor Emeritus} \Email{ bertossi@scs.carleton.ca}\\
 \addr Carleton University, Ottawa, Canada.   %
}

\usepackage{color}
\usepackage{hhline}
\usepackage{tabu}
\usepackage{caption}
\usepackage{multirow} 

\newcommand{\ignore}[1]{}

\newcommand{\boxtheorem}{\hfill $\blacksquare$\\}

\newcommand{\msf}[1]{\mbox{\sf #1}}

\ignore{++
\makeatletter
\newcommand{\pushright}[1]{\ifmeasuring@#1\else\omit\hfill$\displaystyle#1$\fi\ignorespaces}
\makeatother
++}

\newcommand{\red}[1]{\textcolor{red}{#1}}
\newcommand{\blue}[1]{\textcolor{blue}{#1}}

\newcommand{\ces}{CES(I,U($\frac{1}{2}$))}
\newcommand{\In}{\msf{\footnotesize \ in}}
\newcommand{\Out}{\msf{\footnotesize \ out}}
\newcommand{\nit}[1]{{\it #1}}

\newcommand{\mc}[1]{\mathcal{ #1}}
\newcommand{\phill}{\phantom{po}  \hfill}

\newcommand{\comlb}[1]{{\vspace{2mm}\noindent \bf \red{COMM(LEO):}}~ #1   \hfill {\bf
		\red{END.}}\\}
\newcommand{\comfa}[1]{{\vspace{2mm}\noindent \bf \blue{COMM(FEL):}}~ #1   \hfill {\bf
	\blue{END.}}\\}

\newcommand{\mbb}[1]{\mathbb{ #1}}

\ignore{++
\newcounter{example-counter}
\newcounter{remark-counter}
\renewenvironment{example}%
{\vskip \abovedisplayskip \refstepcounter{example-counter}%
\noindent {\bf Example \arabic{example-counter}.}}%
\renewenvironment{remark}%
{\vskip \abovedisplayskip \refstepcounter{remark-counter}%
\noindent {\bf Remark \arabic{remark-counter}.}}%
\setcounter{example-counter}{0}
\setcounter{remark-counter}{0}
++}

\begin{document}
\setcitestyle{authoryear} 

	\maketitle

	\begin{abstract}
	The {\em Causal Effect} (CE) is a numerical measure of  causal influence of variables on observed results. Despite being widely used in many areas, only preliminary attempts have been made to use CE as an attribution score in data management, to measure the causal strength of tuples for query answering in databases. In this work, we introduce, generalize and investigate the so-called {\em Causal-Effect Score} in the context of classical and probabilistic databases.
	\end{abstract}


	\section{Introduction}\label{sec:introduction}

Users of a database management system could expect {\em explanations}, for example, for the answers to a query, for a violation of an integrity constraint, etc. So as in  AI and machine-learning,  explanations may come in different forms, in particular, as an {\em attribution score}, that is, as a quantitative degree of relevance of a piece of data in a database (DB) in relation to the observed output (or for the feature values in input entities in the case of AI and ML). \ 
 This work concentrates  on {\em local attribution scores} as explanations in DBs, that is, for a particular query answer, and at the DB tuple level.

Several scores have been proposed and investigated in DBs. \ {\em Responsibility} \citep{chockler} is based on {\em actual causality} and  {\em counterfactual interventions} \citep{HP05,halpern15,H16}. The latter can be seen as hypothetical changes performed on a (causal) model, to identify other changes. By doing so, one can detect cause-effect relationships.
\  In the context of actual causality, {\em Responsibility} (RESP) has been applied in DBs to quantify the relevance of individual tuples, and attribute values in them, for a query result \citep{QA_Causality,QA_CausalityII,tocs,flairs17}.
\ The  {\em Shapley Value} of {\em Coalition Game Theory} \citep{shapley1953original,roth1988shapley}, as a measure of contribution of individual players to a shared wealth or game function, has been applied as an explanation score in DBs, with tuples acting as  players, and the query (Boolean or aggregate) as the game function \citep{Shapley_Tuple_Bertossi,sigRec21,benny22,sigRec23}. \  The Shapley Value is as the only measure that satisfies a given set of desired properties (or {\em axioms}) \citep{roth1988shapley}. A related game-theoretic measure, the {\em Banzhaf Power Index} (BPI)  \citep{banzhaf_value}, has been applied in DBs as the {\em Banzhaf Score}
\citep{Shapley_Tuple_Bertossi,deutchBanzhaf}.

 The {\em Causal-Effect Score} (CES) can be traced back to {\em causality in observational studies}  \citep{rubin,holland}, where one usually cannot predefine and build control groups, but they have to be recovered from the available data  \citep{gelman,Pearl_CI,roy&salimi23}. The CES is also based on  interventions. 
 \ In \citep{Causal_Effect},
an appropriate form of the CES was first used in data management, when RESP did not provide intuitive results. \ Furthermore,  CES is more appropriate than RESP for explaining  aggregate queries. RESP would only consider, say under  a counterfactual tuple-deletion, if there is a change in the numerical value, no matter how small.  CES would take the amount of change into consideration.

\ In order to apply the CES with a query posed to a DB $D$ \citep{Causal_Effect}, \ $D$ is first transformed  into a {\em tuple-independent probabilistic DB} (TID) $D^p$ \citep{suciu} that {\em independently} and {\em uniformly} assigns  a probability  of $\frac{1}{2}$ to all tuples. The ``counterfactual game" is played on $D^p$.  \ The CES can accommodate both {\em endogenous} and {\em exogenous} tuples, which have also been considered with  RESP and the Shapley Value in DBs \citep{QA_Causality,tocs,Shapley_Tuple_Bertossi}. \ Sometimes we will denote this case of the CES with CES(I,U($\frac{1}{2}$)).

In this work, we propose and investigate
a {\em Generalized Causal-Effect Score} (GCES) in DBs. Generalizing \ces, it can be applied to arbitrary probabilistic DBs.  We introduce the GCES on the basis of {\em counterfactually intervened distributions}. \ The GCES becomes  interesting in several situations, among them, when: \ (a) \ One cannot assume independence or a uniform distribution on tuples, in particular with {\em block-independent} PDBs \citep{suciu}. \ (b) \ Additional domain semantics or domain knowledge can be taken into account, e.g. integrity constraints on the DB (this was investigated for RESP in \citep{flairs17}). ICs can be compiled into an updated joint distribution \citep{adbis23}. \ (c) \ There are explicit correlations among database tuples \citep{deshpande0,deshpande}. \ (d) \ There are quantitative or qualitative stochastic (in)dependencies among attributes provided by a Bayesian or causal network \citep{getoor}. \ (e) \ With crowd-sourced data that come with different
trustworthiness scores or; and, more generally, when data from sources of varying reliability are combined \citep{ratner20}. 

 The GCES can be defined for any query that returns a numerical value ($0$ or $1$ for Boolean query, and a number for a scalar aggregate query). However, we restrict its analysis to monotone queries, which include those most common queries. This is a common practice in database research, as a first step, to gain insight before moving to more general queries. We investigate the data complexity of computing causal-effect scores as explanations for query results, establishing a dichotomy for Boolean Conjunctive Queries (BCQs). \ Furthermore,
taking inspiration from the categorical axiomatization of the Shapley value \citep{roth1988shapley}, we uncover an axiomatic characterization of the GCES as applied to DBs and monotone queries. We compare these properties with those of the Shapley Value and the BPI. \ Notice that, for 
 query explanations in DBs,  the latter coincides with CES(I,U($\frac{1}{2}$))  \ \citep{Shapley_Tuple_Bertossi}.

 This paper is structured as follows. Section \ref{sec:backgrounda} provides background on databases, probabilistic databases, and the Shapley Value and BPI. It also describes related work. Section \ref{sec:CE} introduces the Generalized Causal Effect Score. \   Section \ref{sec:complexity} investigates the  complexity of computing the GCES.  \ Section \ref{sec:ces_props} analyzes the general properties of the GCES, and its axiomatization. \ Section \ref{sec:conclusions} points to ongoing and future research. \ Proofs  can be found  in the Appendix.

\vspace{-3mm}
	\section{Background}
	\label{sec:backgrounda}

\vspace{-1mm}
\paragraph{\bf Relational Databases.} For causality purposes, some of the tuples in a DB $D$ are considered to be {\em endogenous}. They can be subject to causal  (more precisely,  counterfactual) {\em interventions}, in this case, deletions, insertions, or value updates. The other tuples are {\em exogenous}, and are taken as given. They may participate in query answering, but they are not subject to interventions.
 \ $D^\nit{en}, D^\nit{ex}$ denote the subinstances containing the endogenous, resp. exogenous tuples. \ With $\nit{Adom}(D)$ we denote the {\em active domain} of $D$, formed by all the constants appearing in $D$.

   In this work, we will consider {\em Boolean conjunctive queries} (BCQs), unions thereof (UBCQs), which all  take
values $0$ or $1$ on DBs; and also,  numerical aggregations on CQs. \ $D \models \mc{Q}$ denotes that instance $D$ makes $\mc{Q}$ true. $\mc{Q}[D]$ denotes the  answer to query $\mc{Q}$ on instance $D$.  A query $\mc{Q}$ is  {\em monotone} when, for every $D_1 \subseteq D_2$, it holds $\mc{Q}[D_1] \leq \mc{Q}[D_2]$. BCQs and UBCQs are monotone.
   \ All the computational complexity results refer to {\em data complexity}, i.e. in the size of the database.

 \paragraph{\bf Probabilistic Databases.} We recall here what we need about probabilistic databases (PDBs). See \citep{suciu} for a deeper treatment. \ For an initial intuition, we can conceive a PDB as a regular relational DB whose relations have an extra attribute accommodating probability values associated to the corresponding tuples. \
Table \ref{tab:pdbs} shows a PDB of this kind. For example, the first tuple in the relation $R$  indicates that tuple $\tau_1$ belongs to the relation  with probability $p_1$.

\vspace{2mm}

\begin{table}[h]
\begin{center}
{\footnotesize
$\begin{tabu}{c|c|c|c||c|}\hline
R & A & B & C& P \\ \hline
\tau_1& a_1 & b_1& c_1&p_1\\
\tau_2 & a_2& b_2& c_2&p_2\\
\tau_3 & a_3& b_3& c_3&p_3\\
\tau_4 & a_4 & b_4& c_4 &p_4\\
\tau_5& a_5& b_5& c_5&p_5\\
\hhline{~----}
\end{tabu}$}\vspace{-2mm}
\caption{A TID}\label{tab:pdbs}
\end{center}
\end{table}

\vspace{-4mm}
The semantics of a PDB, $D^p$, is a {\em possible world semantics}, in this case a collection $\mc{W}$ of subinstances, $W$, of $D^p$ whose relations, $R_W$, do not have a probabilistic attribute, but a global  probability, $p(R_W)$. Different semantics differ on how the instances $D$ are built and how probabilities are assigned to their relations.

The most common case is that of a {\em tuple-independent} PDBs (TIDs). In  Table \ref{tab:pdbs}, each tuple is (or is not) in the DB independently from the other tuples in the relation  and in other relations. Each tuple has a probability assigned. In a possible world $W \in \mc{W}$, the corresponding (non-probabilistic) relation $R_W$ will contain only some of the tuples in $R$, and the probability associated to $R_W$ is defined by: \ $p(R_W) := \Pi_{_{\tau_i \in R_W}} p_i \ \times \ \Pi_{_{\tau_j \in (R \smallsetminus R_W)}} (1-p_j)$.  For example, if in a possible world $W$ relation $R_W$  contains only tuples $\tau_1$ and $\tau_3$, it  will have the probability $p_1 \times (1-p_2) \times p_3 \times (1-p_4) \times (1-p_5)$. \  Tuple independence beyond the single-relation level leads to the overall probability assigned to a possible world: \ $p(W) := \Pi_{_{R_W}} p(R_W)$.

More generally, for the purpose of this work, and not necessarily in the TID case,  a PDB  $D^p$ can be identified with a discrete {\em probability space} $\langle \mc{W}, p\rangle$, where $\mc{W}$ is the collection of {\em possible worlds} $W$ that are regular relational subinstances of $D^p$,  \ $p$ is defined on $\mc{W}$,\ and  $\sum_{_{W \in \mc{W}}} p(W) =1$. \ignore{ \ When $D$ is clear from the context, we simply write $\mc{W}$ and $p$.} \
Since, $D^p$ is finite, $\mc{W}$ and every world  $W$ in it are also finite.

Given a PDB $D^p$, and a tuple  $\tau$, the probability of $\tau$ being in  $D^p$, is defined by: \vspace{-1mm}
\begin{equation}
P(\tau) \ := \sum_{W \in \mc{W}: \ \tau \in W} p(W). \label{eq:tupProb}\vspace{-1mm}
\end{equation}

\begin{remark}\label{rem:exo} \em \ If there is a partition $(D^\nit{en},D^{\nit{ex}})$ of $D^p$, we assume that, for every $W \in \mc{W}$ with $D^{\nit{ex}} \not \subseteq W$, it holds $p(W) = 0$. \ From this we obtain that, for every $\tau \in D^\nit{ex}$, $P(\tau) = 1$.  \boxtheorem
\end{remark}

\vspace{-8mm} A BCQ on a PDB becomes a Bernoulli random variable, taking values $0$ or $1$ on the outcomes $W \in \mc{W}$. \ There are several query-answering semantics that have been considered in the literature. \  We briefly mention the one that becomes relevant in our work.

 Let $\mc{Q}$ be a query for schema $\mc{S}$.  \   If $\mc{Q}$ is a BCQ, the probability of $\mc{Q}$ (being true) is \
$P(\mc{Q},D^p)  :=
\sum_{W \in \mc{W}: \ W \models \mc{Q}} p(W)$. \
If $\mc{Q}(\bar{x})$ is an open query, and $\bar{a}$, a sequence of constants, \ $P(\bar{a}) := P(\mc{Q}[\bar{a}])$.
\ Under this semantics,  each answer comes with a probability (of being an answer).
\ Notice that for a Boolean query, \
$P(\mc{Q},D^p) = \mbb{E}(\mc{Q})$,
which invites us to define, for an {\em aggregate query} $\mc{Q}$ a probabilistic answer: \
$\mc{Q}[D^p]  := \mbb{E}(\mc{Q}) = \sum_{W\in \mc{W}} p(W) \times \mc{Q}[W]$. \
We will often omit the database when it is clear from the context, simply writing $P(\mc{Q})$.

 \paragraph{\em \bf Shapley Value and Banzhaf Power-Index in Databases.} The Shapley Value \citep{shapley1953original} was used in \citep{Shapley_Tuple_Bertossi}  to quantify the contribution of tuples to a query answer (see also \citep{sigRec21,sigRec23}), as follows: \vspace{-1mm}
    \begin{equation}
    \nit{Shapley}(D,\mc{Q},\tau) = \sum_{S \subseteq (D^\nit{en} \smallsetminus \{\tau\})} \dfrac{|S|! \cdot (|D^\nit{en}| - |S| - 1)!}{|D^\nit{en}|!}  \cdot \Delta(\mc{Q},S,\tau), \label{eq:shap}
    \end{equation}
    with \  $\Delta(\mc{Q},S,\tau) \  := \ \mc{Q}[S \cup D^\nit{ex} \cup \{\tau\}] - \mc{Q}[S \cup D^\nit{ex}]$. \ When $\mc{Q}$ is Boolean,  $\mc{Q}[S]$ is $0$ or $1$. When it is a numerical aggregation, $\mc{Q}[S]$ is the resulting value.
    \ignore{  \begin{equation}\Delta(\mc{Q},S,\tau) \  := \ \mc{Q}[S \cup D^\nit{ex} \cup \{\tau\}] - \mc{Q}[S \cup D^\nit{ex}].\label{eq:delta}
    \end{equation} Here, $\mc{Q}[S]$ is $0$ or $1$ when $\mc{Q}$ is Boolean and false, resp. true in $S$. $\mc{Q}$ can also be a numerical aggregation, and $\mc{Q}[S]$ is its numerical value. }
    \ The {\em Bahnzhaf Power-Index} (BPI) \citep{banzhaf_value} is defined, for queries in DBs, by: \vspace{-2mm}
\begin{equation}
  \nit{BPI}(D,\mc{Q},\tau) \ := \sum_{S \subseteq D^\nit{en} \smallsetminus \{\tau\}} \dfrac{1}{2^{|D^\nit{en}|-1}} \cdot \Delta(\mc{Q},S,\tau).  \label{eq:banzhaf}
    \end{equation}

\vspace{-2mm}
The Shapley  {\em is the only measure} of contribution that satisfies certain desirable properties \citep{shapley1953original,roth1988shapley}. The more relaxed definition of the BPI (permutations are not considered) makes it miss some of the properties of the Shapley value (see Section \ref{sec:ces_props}).

\paragraph{\bf Related Work.} In \citep{Shapley_Tuple_Bertossi}, the  Shapley value and the BPI  were applied and investigated in data management, to quantify the contribution of individual tuples to a query answer (see also \citep{sigRec21,sigRec23}. \ In \citep{bienvenu}, the complexity of the Shapley value in DBs is further explored by establishing a connection with the Generalized Model Counting Problem (the number of subinstances of a given size that satisfy the query). \ In this line of research, \citep{benny22} concentrates on computational aspects of the Shapley value applied to query answering.

In \citep{deutch2}, the {\em rankings} induced by the Shapley values are investigated, rather than on the values themselves. In \citep{deutch3} the interest is in the combination of {\em data provenance} and the Shapley values, for the computation of the latter.
 In \citep{deutchBanzhaf}, research  delves more deeply into experimental results around the  BPI for query answering.
\
In \citep{kara}, the problem of computing the Shapley value of variables in Boolean circuits  is connected with query evaluation in PDBs, obtaining  dichotomy results (similar to those in \citep{dichotomy_UCQ}) for the complexity of Shapley value computation  for query answering in DBs.

Applications of the responsibility score in data management (DM) have been less explored than those of the Shapley value. References for the use of responsibility in  DM are \citep{QA_CausalityII,tocs,flairs17}, among others. In all those papers, semantic and computational problems were addressed. \ Close to responsibility as used in DM, we find the notion of {\em resilience}, whose computational aspects have been investigated \citep{wolfgang}.

The use of the {\em causal effect} in DM has been explored, and only superficially and in its simplest form, i.e. the CES(I,U($\frac{1}{2}$)), in  \citep{Causal_Effect}. The connection with the BPI was first established in  \citep{Shapley_Tuple_Bertossi}.  The complexity of CES(I,U($\frac{1}{2}$)) has been partially and indirectly studied in \citep{Shapley_Tuple_Bertossi}, through its connection to the BPI.

Closer in spirit to our work, in that a probabilistic setting is assumed at the start, \citep{senellart} investigate the expected value of the Shapley value on tuple-independent PDBs.

\vspace{-3mm}
    \section{The Causal-Effect Score in Databases}
    \label{sec:CE}

In data management, we usually want to explain {\em why} a given query becomes true or returns a particular answer. In this work, we want to provide {\em explanations at the tuple-level}. \ In   \citep{Causal_Effect}, it was shown that the {\em Causal-Effect Score} (in its CES(I,U($\frac{1}{2}$)) form)   can be sensibly applied as an explanation score that  reflects the causal strength of a tuple for the query answer.

\subsection{Interventional Distributions on PDBs}\label{sec:inter}

The causal effect relies on  {\em interventions}, which in DBs become hypothetical insertions or deletions of tuples. If we start with a PDB $D^p$, with a given distribution $p$, an intervention induces a new, {\em interventional distribution} (see Definition \ref{def:DO}). \ 
	Interventions will be denoted with  $\nit{do}(\tau \In)$ and $\nit{do}(\tau \Out)$, with the intuitive meaning that tuple $\tau$ is {\em  made true}, i.e. it is inserted into $D^p$ (if it is not already in it). Similarly,  $\nit{do}(\tau \Out)$ means that $\tau$ is {\em made false}, i.e. removed from $D^p$. Interventions are applied only with endogenous tuples, which have an initial probability of being (true) in $D^p$. Interventions are applied  to detect if making a tuple true or false affects a query answer. They can also be applied with sets of tuples $\mc{T}$, e.g. $\nit{do}(\mc{T} \In)$.\footnote{It is common to apply interventions on variables of a causal model \citep{Pearl_CI}. In the case of databases, to the {\em lineage of the query}, that can act as the model, with  random propositional variables $X_\tau$, which are made {\em true} or {\em false} via $\nit{do}(X_\tau =1)$ or  $\nit{do}(X_\tau =0)$ \ \citep{Causal_Effect}. It should be noted that the notion of causality considered here does not attempt to model some kind of causal
structure in the ``real world”, but is based on the causal structure between the input and output of a database query
inside the computer.}

 We will use expressions of the form  $P(\mc{Q}=1~|~\nit{do}(\tau \In))$, etc., where $\mc{Q}$ is a Boolean query, and $P$ is the intervened distribution on the query range  associated to a PDB $D^p$. Intuitively, it means ``the probability of the query being true given that tuple $\tau$ is made true". \ This common notation may be misleading,  suggesting a conditional probability, which strictly speaking is not. 
 
 More precisely, in Definition \ref{def:DO}, we start with a distribution $p$ on  the class of possible worlds $\mc{W}$ associated to $D^p$, and we create the intervened distributions $p^{+\tau}(W) := p(W~|~\nit{do}(\tau \In))$ and $p^{-\tau}(W) := p(W~|~\nit{do}(\tau \Out))$ on a collection of possible worlds. They  can be seen as  modifications of  the original distribution $p$.\footnote{The starting set $\mc{W}$ of possible world contains subinstances of instance $D^p$ whose probabilities add up to $1$. However, it does not have to contain all subinstances (but commonly it does). Similarly, for the collection of possible worlds after an intervention.} 

\vspace{-1mm}
 \begin{definition}\label{def:DO} \em Given a PDB $D^p=\langle \mc{W}, p\rangle$, and a tuple $\tau$: \\ 
 (a) The positive  intervention with $\tau$ on $D^p$ is the PDB: \ $D^p(\nit{do}(\tau \In)) := \langle \mc{W}^{+\tau},p^{+\tau}\rangle$, with $\mc{W}^{+\tau} := \{W \cup \{\tau\}~|~W \in \mc{W}\}$; \ and, for each $W^\prime \in \mc{W}^{\tau +}$\!\!, \    $p^{+\tau}(W^\prime) := \sum_{W \cup \{\tau\} = W^\prime}  p(W)$. \\
 (b) The negative  intervention with $\tau$ on $D^p$ is the PDB: $D^p(\nit{do}(\tau \Out)) := \langle \mc{W}^{-\tau},p^{-\tau}\rangle$, with
  $\mc{W}^{-\tau} := \{W \smallsetminus \{\tau\}~|~W \in \mc{W}\}$; \ and, for each $W^\prime \in \mc{W}^{\tau -}$\!\!, \
  $p^{-\tau}(W^\prime):= \sum_{W \smallsetminus \{\tau\} = W^\prime}  p(W)$. \newline 
   (c) \  For a Boolean query $\mc{Q}$, its intervened probabilities are:

\begin{eqnarray*}P(\mc{Q}=1~|~\nit{do}(\tau \mbox{\phantom{o}}\In)) &:=&
p^{+\tau}( \{W^\prime \in \mc{W}^{+\tau}~|~W^\prime \models \mc{Q}\} ), \\ 
P(\mc{Q}=0~|~\nit{do}(\tau \mbox{\phantom{o}}\In)) &:=&
p^{+\tau}( \{W^\prime \in \mc{W}^{+\tau}~|~W^\prime \not \models \mc{Q}\} ), \\ 
P(\mc{Q}=1~|~\nit{do}(\tau \Out)) &:=&
p^{-\tau}( \{W^\prime \in \mc{W}^{-\tau}~|~W^\prime \models \mc{Q}\} ), \\ 
~~~P(\mc{Q}=0~|~\nit{do}(\tau \Out)) &:=&
p^{-\tau}( \{W^\prime \in \mc{W}^{-\tau}~|~W^\prime \not \models \mc{Q}\} ). 
\end{eqnarray*}
\end{definition}

\vspace{-0.7cm}\phill $\blacksquare$

\vspace{2mm}
It holds:  
$p^{+\tau}( \{W^\prime \in \mc{W}^{+\tau}~|~W^\prime \models \mc{Q}\})  
= \sum_{W \in \mc{W}: \ W\cup \{\tau\} \models \mc{Q}} p(W)$  (similarly for other cases).    In case (c), $P$ is the probability induced by, e.g., $p^{+\tau}$ on the range $\{0,1\}$ of the query.
\ Equivalently, the intervened PDBs and query answering can be expressed as follows:

 \vspace{1mm} \noindent  (a) \ $D^p(\nit{do}(\tau \In)) \ := \ \langle \mc{W},p^{+\tau}\rangle$, such that, for $W \in \mc{W}$: 
 $p^{+\tau}(W)  \ := \sum_{W^\prime \in \mc{W} \ : \ W^\prime \cup \{\tau\} = W}  p(W^\prime)$.

 \vspace{1mm} \noindent (b) \ $D^p(\nit{do}(\tau \Out)) \ := \ \langle \mc{W},p^{-\tau}\rangle$, such that,  for $W \in \mc{W}$: $p^{-\tau}(W) \ := \sum_{W^\prime \in \mc{W} \ : \ W^\prime \smallsetminus \{\tau\} = W}  p(W^\prime)$.

\vspace{1mm}
  \noindent
 (c) \ For a BQ $\mc{Q}$, and $v \in \{0,1\}$: \vspace{1mm}
 
 \hspace*{7mm}$P(\mc{Q}=v~|~\nit{do}(\tau \mbox{\phantom{o}}\In)) \ := \
p^{+\tau}( \{W \in \mc{W}~|~\mc{Q}[W]=v\} )$, \ \ and\\
\hspace*{13mm}$
P(\mc{Q}=v~|~\nit{do}(\tau \Out)) \ := \
p^{-\tau}( \{W \in \mc{W}~|~ \mc{Q}[W] =v\})$.

\begin{remark}\label{rem:tid} \em (a) \ If we consider $\tau$ as a ground atomic query, as a particular case of
Definition \ref{def:DO}(c), we obtain: \ $P(\tau~|~\nit{do}(\tau \In)) := P(\tau = 1~|~\nit{do}(\tau \In)) = \sum_{W \in \mc{W}, \ \tau \in W\cup \{\tau\}} p(W) = 1$. \ Similarly, \
$P(\tau~|~\nit{do}(\tau \Out))  = 0$, \ which captures the original intuition.

\vspace{1mm}\noindent  (b)
 \ In Definition \ref{def:DO}(c), it holds: \ $P(\mc{Q}=1~|~\nit{do}(\tau \mbox{\phantom{o}}\In)) \
= \ \sum_{W \in \mc{W}, \ W\cup \{\tau\} \models \mc{Q}} p(W)$. \ Similarly for the other cases.

\vspace{1mm} \noindent
(c) \ When \ $\tau \notin W \in \mc{W}, \ p^{+\tau}(W) = 0$; \ and, when $\tau \in W \in \mc{W}, \ p^{-\tau}(W) = 0$.

\vspace{1mm}\noindent  (d)
\ For a TID $D^p$ and different tuples $\tau,\tau^\prime \in D^\nit{en}$, it holds:
    $P(\tau^\prime ~|~\nit{do}(\tau \In))  = P(\tau^\prime ~|~\nit{do}(\tau \Out)) = p(\{\tau^\prime\})$,
that is, an intervention $\nit{do}(\tau \In)$ ($\nit{do}(\tau \Out)$, resp.) on a TID translates in changing the probability of $\tau$ to 1 (0, resp.), leaving all other probabilities unchanged. \ Furthermore, an intervened TID becomes a TID. \boxtheorem
\end{remark}

\vspace{-1cm}

\begin{table}[h]
   \begin{center}
 {\scriptsize
 $\begin{tabu}{l|c|c|}
				\hline
				E~  & ~~A~~ & ~~B~~ \\\hline
				\tau_1 & a & b\\
				\tau_2& a & c\\
				\tau_3& c & b\\
				\tau_4& a & d\\
				\tau_5& d & e\\
				\tau_6& e & b\\ \cline{2-3}
			\end{tabu}$}~~~~~~~~~~~~~~~~~
 {\footnotesize
 $\begin{tabu}{l|c|c|}
				\hline
				\mc{W}~  & ~~\mbox{extension}~ & ~~p~~ \\\hline
				W_1 & \{\tau_1,\tau_3,\tau_4,\tau_6\} & 0.20\\
				W_2&  \{\tau_1, \tau_2, \tau_3\}& 0.25\\
				W_3& \{\tau_2, \tau_3, \tau_6\}& 0.15\\
				W_4& \{\tau_2, \tau_6\}&  0.40\\ \cline{2-3}
			\end{tabu}$}
\end{center}
\vspace{-4mm}\caption{ \ An instance and its associated PDB}\label{tab:classic}\vspace{-2mm}
\end{table}

\ignore{+++ Old example in submitted version, replaced by its full version in ICDT26 submission

\begin{example} \em\label{ex_paths1}  \
		Let $D$ consist of relation $E$ in Table \ref{tab:classic}, with its  associated non-TID $D^p$. All  tuples are endogenous.  \ For  $W \in (\mc{W} \smallsetminus \{W_1,W_2,W_3,W_4\})$, \ $p(W) := 0$. For example, \ $p^{+\tau_3}(W_3) =  p(W_3) + p(W_4)= 0.55$; \ and \
$p^{-\tau_3}(W_4) = p(W_3) + p(W_4) =  0.55$.\footnote{This example is developed in full detail in Example \ref{ex_paths1App} in Appendix \ref{sec:examples}.} \boxtheorem
\end{example}
+++}

\vspace{-0.5cm}\begin{example} \em \label{ex_paths1App}  \ 
		Let $D$ consist of relation $E$ in Table \ref{tab:classic}, with all  tuples endogenous; and  let us define a PDB $D^p$ as in the Table. \   For  $W \in (\mc{W} \smallsetminus \{W_1,W_2,W_3,W_4\})$, \ $p(W) := 0$.

It holds: \ $P(\tau_1) = 0.2 + 0.25 = 0.45$, \ $P(\tau_2) = 0.8, \ P(\tau_3) = 0.8, \ P(\tau_4) = 0.2, \ P(\tau_5) = 0, \ P(\tau_6) = 0.75$. \
Notice that  \ $P(W_1) = 0.20 \neq 0 = P(\tau_1) \times (1-P(\tau_2)) \times P(\tau_3) \times P(\tau_4) \times (1-P(\tau_5)) \times P(\tau_6)$, which shows that $D^p$ is not a TID.
\ Let us compute, e.g. the intervened distribution \ $p^{+\tau_3}(\cdot) = p(\cdot ~|~\nit{do}(\tau_3 \In))$:

\vspace{-5mm}\begin{eqnarray*}
p^{+\tau_3}(W_1) &=& \sum_{W^\prime \in \mc{W}: \  W^\prime \cup \{\tau_3\} = W_1} p(W^\prime) = p(W_1) = 0.20. \\
p^{+\tau_3}(W_2) &=& \sum_{W^\prime \in \mc{W}: \  W^\prime \cup \{\tau_3\} = W_2} p(W^\prime) = p(W_2) = 0.25.
\end{eqnarray*}

\begin{eqnarray*}
p^{+\tau_3}(W_3) &=& \sum_{W^\prime \in \mc{W}: \ W^\prime \cup \{\tau_3\} = W_3} p(W^\prime) = p(W_3) + p(W_4)= 0.55.\\
p^{+\tau_3}(W_4) &=& \sum_{W^\prime \in \mc{W}: \ W^\prime \cup \{\tau_3\} = W_4} p(W^\prime) = 0.
\end{eqnarray*}
For any other $W \in \mc{W}$, \ $p^{+\tau_3}(W) = 0$. \ Now, the intervened distribution \ $p^{-\tau_3}(\cdot) = p(\cdot ~|~\nit{do}(\tau_3 \Out))$.
\begin{eqnarray*}
p^{-\tau_3}(W_i) &=& \sum_{W^\prime \in \mc{W}: \  W^\prime \smallsetminus \{\tau_3\} = W_i} p(W^\prime) = 0, \ \ \  i=,\ldots,3. \\
p^{-\tau_3}(W_4) &=& \sum_{W^\prime \in \mc{W}: \ W^\prime \smallsetminus \{\tau_3\} = W_4} p(W^\prime) = p(W_3) + p(W_4) =  0.55.\\
p^{-\tau_3}(\{\tau_1,\tau_4,\tau_6\}) &=& \sum_{W^\prime \in \mc{W}: \  W^\prime \smallsetminus \{\tau_3\} = \{\tau_1,\tau_4,\tau_6\}} p(W^\prime) = p(W_1) = 0.20.\\
p^{-\tau_3}(\{\tau_1,\tau_2\}) &=& \sum_{W^\prime \in \mc{W}: \  W^\prime \smallsetminus \{\tau_3\} = \{\tau_1,\tau_2\}} p(W^\prime) = p(W_2) = 0.25.
\end{eqnarray*}
For any other $W \in \mc{W}, \ p^{-\tau_3}(W) = 0$.  \boxtheorem
\end{example}

The definitions for query answering can be easily modified for a scalar aggregate query $\mc{Q}$, which also becomes a random variable over a PDB. Similarly, and ``componentwise", for aggregate queries with group-by.

\subsection{The Generalized Causal Effect Score}
\label{sec:gce}

In \citep{Causal_Effect}, in order to apply the \ces, the original DB was converted into a uniform TID. \ The \ces \ can be generalized by considering an arbitrary PDB  $D^p$.

    \begin{definition} \label{def:gce} \em
        Let $D^p = \langle \mc{W},$ $p \rangle$ be a PDB , and
        $\mc{Q}$ a Boolean or scalar aggregate query. \ The {\em generalized causal effect score} \ (GCES) of $\mc{T} \subseteq D^{\nit{en}}$ on $\mc{Q}$ is: \vspace{3mm}
        
        \hspace*{2.5cm}$\nit{CE}(D^p,\mc{Q},\mc{T}) \ := \ \mathbb{E}(\mc{Q}~|~ \nit{do}(\mc{T} \In)) - \mathbb{E}(\mc{Q}~|~\nit{do}(\mc{T} \Out))$.\boxtheorem  
    \end{definition}

\vspace{-3mm}This definition relies on the probability distribution of $D^p$. When $D^p$ is a TID associated to a regular relational DB $D$, we write  $\nit{CE}^I\!(D,\mc{Q},\mc{T})$. \ \ces \ is a particular case where endogenous tuples have a uniform distribution $U(\frac{1}{2})$ (see Remark \ref{rem:exo}). \   In this case we use the notation  $\nit{CE}^\nit{\!UI}\!(D,\mc{Q},\mc{T})$ (omitting the $\frac{1}{2}$).

The role of exogenous tuples is twofold. First, they are excluded from counterfactual interventions; and, secondly, they influence the probabilities used in the expected values by contributing with probability $1$.  \ There are no technical difficulties in defining causal-effects for exogenous tuples. However, we leave them outside the scope of explanations for query answers.

\begin{remark}  \label{rem:gcs} \ {\bf Special cases and notation.} \em \ (a) \ For a Boolean query $\mc{Q}$:  \
$\nit{CE}(D^p,\mc{Q},\mc{T}) = P(\mc{Q} = 1~|~\nit{do}(\mc{T} \In))- P(\mc{Q} = 1~|~\nit{do}(\mc{T} \Out))$. \ (b) \ $\nit{CE}(D^p,\mc{Q},\tau)$ denotes the GCES for a single endogenous tuple $\tau$.   \boxtheorem
\end{remark}

\vspace{-4mm}In \citep{Causal_Effect} it was shown that, for a
	 MBQ $\mc{Q}$,  and $\tau \in D^\nit{en}$,  $\tau$ is an \emph{actual cause} \citep{HP05} for $\mc{Q}$ in $D$ iff $\nit{CE}^{\!\nit{UI}\!}(D,\mc{Q},\tau) > 0$. \  In \citep{Shapley_Tuple_Bertossi}, it was shown that the CES coincides with the BPI: \ 
 $\nit{CE}^{\!\nit{UI}\!}(D,\mc{Q},\tau) = \nit{BPI}(D,\mc{Q},\tau)$.

\begin{example} \label{ex_paths} \em \
		Let $D$ be a database instance with the relations $E$ and $S$ here below \ignore{in Table \ref{tab:ce_ex}} \  (used in  \citep{Causal_Effect}), with all their tuples endogenous. \ Let us  build a {\em uniform} TID by defining: \   $p(\tau) := \frac{1}{2}$ for every tuple $\tau$. \ Consider the  Boolean query $\mc{Q}_1$ asking if there exists a path from $a$ to $b$ according to relation $E$. It can be expressed in Datalog (or a UCQs for a fixed instance), which makes it monotone.

\begin{center} \phantom{o}
{\scriptsize
 $\begin{tabu}{l|c|c|}
				\hline
				E~  & ~~A~~ & ~~B~~ \\\hline
				\tau_1 & a & b\\
				\tau_2& a & c\\
				\tau_3& c & b\\
				\tau_4& a & d\\
				\tau_5& d & e\\
				\tau_6& e & b\\ \cline{2-3}
			\end{tabu}$~~~~~~~~~~
            $\begin{tabu}{l|c|c|}
				\hline
				S   & ~~A~~ & ~~C~~\\
				\hline
				\tau_7 & a & 1\\
				\tau_8 & a & 2\\
				\tau_9 & b & 0\\
				\tau_{10} & a & 3\\
				\tau_{11} & b & 1\\
				\tau_{12} & b & 10\\
				\cline{2-3}
			\end{tabu}$}
\vspace{-2mm}
\end{center}

 
 \vspace{-2mm} It holds: $\nit{CE}^\nit{\!UI\!}(D,\mc{Q}_1,\tau_1) =
   0.65625$, $\nit{CE}^\nit{\!UI\!}(D,\mc{Q}_1,\tau_2) = \nit{CE}^\nit{\!UI\!}(D,\mc{Q}_1,\tau_3) =$  $0.21875$, and
   $\nit{CE}^\nit{\!UI\!}(D,\mc{Q}_1,$ $\tau_4) = \nit{CE}^\nit{\!UI\!}(D,\mc{Q}_1,\tau_5) = \nit{CE}^\nit{\!UI\!}(D,\mc{Q}_1,\tau_6) = 0.09375$. \  As noticed in \citep{Causal_Effect}, these scores significantly differ from the responsibility scores, which are all  $\frac{1}{3}$, despite the fact that they make the query true through paths of different lengths.

  Consider now the  scalar aggregate query $\mc{Q}$ defined by: \  $\nit{Ans}^{\!\mc{Q}\!}(\nit{sum}(y)) \leftarrow S(x,y)$. \ Its answer is $17$. \ When $\nit{Dom}(B) \subseteq \mbb{R}^+$,  this is a monotone query. \
  \ces \  for a tuple $\tau \in S$ is computed using Definition \ref{def:gce}. We need to compute the expected value of the query when intervening the tuple $\tau$. Denote with $\tau[C]$ the restriction of a tuple $\tau$ to attribute $C$; in this case, the numerical value. Consider that, for any $\tau^{\prime} \in D$ with $\tau \neq \tau^{\prime}$,  the probability of $\tau^\prime$ stays the same when intervening $\tau$, i.e. $P(\tau^{\prime}~|~ do(\tau \In)) = P(\tau^{\prime}~|~ do(\tau \Out)) = p(\{\tau^\prime\})$, and for $\tau$, it holds $P(\tau~|~ do(\tau \In)) = 1$ and $P(\tau~|~ do(\tau \Out)) = 0$. Now, and since the aggregate query $\mc{Q}$ is just adding the value of the attribute $C$ from each tuple, the CES becomes: \ $\nit{CE}^\nit{\!UI\!}(D,\mc{Q},\tau_7) = \mbb{E}(\mc{Q}~|~ \nit{do}(\tau_7 \In)) - \mbb{E}(\mc{Q}~|~ \nit{do}(\tau_7 \Out)) = \tau_7[C] \ = \ 1$, a  result in line with the intuition: the average expected contribution of  tuple $\tau_7$ for a query that is adding up all tuples in the relation would be the attribute value of the tuple itself, $\tau_7[C]$.

  Let us now define a PDB $D^p$ (restricted to relation $E$)  by: \ For $W_1 = \{\tau_1, \tau_3, \tau_4, \tau_6\}, \linebreak p(W_1) := 0.20$; \
for $W_2 = \{\tau_1, \tau_2, \tau_3\}, \ p(W_2) := 0.25$; \ for $W_3 = \{\tau_2, \tau_3, \tau_6\}, \linebreak p(W_3) := 0.15$; \ for $W_4 = \{\tau_2, \tau_6\}, \ p(W_4) := 0.40$; and for any other $W \subseteq E$, \ $p(W) := 0$. \ Here, we are starting with a probability distribution over the possible worlds.

According to (\ref{eq:tupProb}): \ $P(\tau_1) = 0.2 + 0.25 = 0.45$, \ $P(\tau_2) = 0.8, \ P(\tau_3) = 0.8, \ P(\tau_4) = 0.2, \ P(\tau_5) = 0, \ P(\tau_6) = 0.75$. \
If we compute the probabilities of possible worlds using these tuple probabilities assuming independence, we obtain, e.g. for $W_1$: \ $P^\prime(W_1) := 0.2 \times (1-0.8) \times 0.8 \times 0.2 \times (1-0) \times 0.75 = 0.00288$, showing that $D^p$ we started with is not a TID.

We can compute the causal effect of $\tau_3$ without explicitly appealing to the intervened worlds and distributions. It differs from $\nit{CE}^{\!\nit{UI}\!}(D,\mc{Q}_1,\tau_3)$: \ \ (see Remark \ref{rem:gcs}(a))

\vspace{1mm}\noindent $\nit{CE}(D^p,\mc{Q}_1,\tau_3) 
= P(\mc{Q}_1=1~|~\nit{do}(\tau_3  \In))  - P(\mc{Q}_1=1~|~\nit{do}(\tau_3  \Out))
= \sum_{W \in \mc{W}, \ W\cup \{\tau_3\} \models \mc{Q}}  p(W) \ \  - \sum_{W \in \mc{W}, \ W\smallsetminus \{\tau_3\} \models \mc{Q}}  p(W)
= (p(W_1) + p(W_2) + p(W_3) + p(W_4)) - (p(W_1) + p(W_2)) = p(W_3) + p(W_4) = 0.55$.

\vspace{1mm}
For illustration, let us consider the intervened distributions $P(\cdot ~|~\nit{do}(\tau_3 \In))$ and  $P(\cdot ~|~\nit{do}(\tau_3 \Out))$ for the computation of the same causal effect: \  $\mc{W}^{+\tau_3}=\{W_1^{\prime},W_2^{\prime},W_3^{\prime}\}$, with $W_1^{\prime} = W_1$, $W_2^{\prime} = W_2$ and $W_3^{\prime} = \{\tau_2, \tau_3, \tau_6\}$; and  
 \  $\mc{W}^{-\tau_3}=\{W_1^{\star},W_2^{\star},W_3^{\star}\}$ with $W_1^\star = W_1$, $W_2^\star = W_2$ and $W^\star_3 = \{\tau_2, \tau_6\}$. 
Notice that $p(W_1) = p^{+\tau_3}(W_1^\prime) = p^{-\tau_3}(W_1^\star)$ and $p(W_2) = p^{+\tau_3}(W_2^\prime) = p^{-\tau_3}(W_2^\star)$. Also, $W_3 \cup \{\tau_3\} = W_4 \cup \{\tau_3\} = W_3^{\prime}$ and $W_3 \!\smallsetminus\! \{\tau_3\} = W_4 \!\smallsetminus\! \{\tau_3\} = W^\star_3$. \ Then,  $p^{+\tau_3}(W_3^{\prime}) = p^{-\tau_3}(W^\star_3) = p(W_3) + p(W_4) = 0.55$.  
\vspace{1mm}

\noindent $\nit{CE}(D^p,\mc{Q}_1,\tau_3) = \mathbb{E}(\mc{Q}_1~|~\nit{do}(\tau_3  \In)) - \mathbb{E}(\mc{Q}_1~|~\nit{do}(\tau_3 \Out))
= \sum_{W^{\prime} \in \mc{W}^{+\tau_3}, \ W^{\prime} \models \mc{Q}_1}  p^{+\tau_3}(W^{\prime})   - \sum_{W^{\star} \in \mc{W}^{-\tau_3}, \ W^{\star} \models \mc{Q}_1} p^{-\tau_3}(W^{\star})
= \left( p^{+\tau_3}(W_1^\prime) + p^{+\tau_3}(W_2^\prime) + p^{+\tau_3}(W_3^\prime)\right) - (p^{-\tau_3}(W_1^\star) \ +$ \linebreak $p^{-\tau_3}(W_2^\star))
\ = \ p^{+\tau_3}(W_3^\prime) \ = \ p(W_3) + p(W_4) = 0.55$. \boxtheorem
\end{example}

\vspace{-9mm}
\section{Computational Complexity of  CES}
    \label{sec:complexity}

     In this section we investigate the complexity of computing the CES for an arbitrary TID, i.e.  $\nit{CE}^I\!(D,\mc{Q},\tau)$, and BCQs \ (see Remark \ref{rem:gcs}).
    \   We obtain a dichotomy result similar to  that  for query answering on TIDs \citep{suciu}. For its formulation, we need some preliminary notions.

     For a BCQ $\mc{Q}$, $\nit{Atoms}(\mc{Q})$ denotes the set of all atoms in $\mc{Q}$, i.e. its conjuncts. \ For a variable $v$ in $\mc{Q}$,  $\nit{Atoms}(v)$ denotes the set of atoms of $\mc{Q}$ where $v$ appears. \ A BCQ $\mc{Q}$ is \emph{hierarchical} if, for any two variables $x,y$ in $\mc{Q}$, one of the following holds: (a) $\nit{Atoms}(x) \subseteq \nit{Atoms}(y)$, (b) $\nit{Atoms}(y) \subseteq \nit{Atoms}(x)$ or (c) $\nit{Atoms}(x) \cap \nit{Atoms}(y) = \emptyset$. \ Otherwise, the query is called \emph{non-hierarchical} \citep{suciu}. \ The evaluation of a BCQ without self-joins on  a TID $D$: \ (a) Can be done in polynomial-time when $\mc{Q}$ is hierarchical, \ and (b) Is  $\#P$-hard when $\mc{Q}$ is non-hierarchical. \ All these results are in data complexity    \citep{suciu}. \ Deciding if a BCQ is hierarchical or not can be done with an efficient syntactic test.

     \begin{definition} \em \label{def:component}
	(a)  Let $\mc{Q}$ be a BCQ; \  $\mc{G}$ the undirected graph with nodes in $\nit{Atoms}(\mc{Q})$ and edges of the form $\{A_i,A_j\}$, where $A_1$ and $A_2$ share at least one variable; and $\mc{C}(\mc{Q})$ the sets of connected components in $\mc{G}$. Accordingly, two different sets of atoms in $\mc{C}(\mc{Q})$ do not share  variables. \ (b) \ For an instance $D$ compatible with the schema of $\mc{Q}$, the {\em fresh expansion of} $D$ {\em via} $\mc{Q}$ is the instance $D \cup \mc{T}$, where $\mc{T}$ contains  exactly one new tuple $\tau^U$ for each $U \in \nit{Atoms}(\mc{Q})$, with the same predicate of $U$ and with the variables inherited from $\mc{Q}$ all replaced simultaneously by new  constants not appearing in $\nit{Adom}(D)$.  \ Now, a {\em selection function} $\sigma$ assigns, to each  $C \in \mc{C}(\mc{Q})$, exactly one  $\tau^U \in \mc{T}$, with $U \in C$.\boxtheorem
\end{definition}

\vspace{-6mm}\begin{remark} \label{rem:ind_probability} \em
   For a TID $D^p$, and a BCQ $\mc{Q}$, it holds: \ 
    $P(\mc{Q},D^p) = \prod_{C_i \in \mc{C}(\mc{Q})} P(\bar{\exists}\mc{Q}_i,D)$, where $\bar{\exists}\mc{Q}_i$ is the existential closure of the conjunction of the atoms in component $C_i$. 
  \ For example, \ for the query $\mc{Q}\!: \exists x \exists y(R_1(x) \wedge R_2(x) \wedge R_3(y))$, \ $C_1 = \{R_1(x), R_2(x)\}, \ C_2 = \{R_3(y)\}$; $\mc{Q}_1$ and $\mc{Q}_2$ are  $(R_1(x) \wedge R_2(x))$ and $R_3(y)$, resp.; and  $P(\mc{Q},D^p) = P(\exists x(R_1(x) \wedge R_2(x)),D^p) \times P(\exists y R_3(y), D^p)$. \boxtheorem
\end{remark}

\vspace{-8mm}
\begin{proposition} \em \label{prop:connection_qep_gces} \ Let $\mc{Q}$ be a BCQ, \ $D^p = \langle \mc{W}, p \rangle$ a TID associated to relational instance $D$, and \ $D^\prime := D \cup \mc{T}$ a fresh expansion of $D$ via $\mc{Q}$. A TID  $(D^\prime)^p = \langle \mc{W}^\prime,p^\prime\rangle$, with $\mc{W}^\prime = \mc{P}(D^\prime)$, the power set of $D^\prime$, and $\mc{T} \subseteq (D^\prime)^\nit{en}$, can be constructed in constant time, in such a way that, for every selection function $\sigma$: \vspace{-1mm}
    \begin{equation} 
   P(\mc{Q},D^p) \ = \ \prod_{C \in \mc{C}(\mc{Q})} (1 - \nit{CE}^{I}(D,\mc{Q},\sigma(C))). \label{eq:prod}
   \end{equation}
 
 \vspace{-5mm}\boxtheorem
    \end{proposition}

\vspace{-9mm} \begin{remark} \em \label{rem:sigmas} \ (a) \ Notice that $P(\mc{Q},D^p)$ can be obtained using any selection function. That is, there are different but coincident representations of this probability. \ (b) The  number of  factors in the product is $|\mc{C}(\mc{Q})|$, and depends only on the query.  \boxtheorem
\end{remark}

\vspace{-6mm}The following example illustrates Definition \ref{def:component} and Proposition \ref{prop:connection_qep_gces}. \ignore{The latter is used to prove Theorem \ref{theo:complCQ}. }

          \begin{example} \ \em Consider the TID $D^p = \langle \mc{W},p \rangle$ here below\ignore{in Table \ref{tab:TID1-CQ}}, and the BCQ  $\mc{Q}\!:\! \exists x \exists y(R_1(x,y)\land R_2(y) \land R_3(z))$.  
          
\vspace*{-5mm}
\begin{center}
{\scriptsize
$\begin{tabu}{l|c|c||c|}
                \hline
                R_1~  & ~~A~~ & ~~B~~ & P  \\
                \hline
                \tau_1 & a & a & 0.9 \\
                \tau_2& b & b & 0.3 \\
                \tau_3& c & b & 0.8 \\
                \cline{2-4}
            \end{tabu}$~~~~~~~~
            $\begin{tabu}{l|c||c|}
                \hline
                R_2 & ~~A~~ & P\\
                \hline
                \tau_4 & b & 0.5\\
                \cline{2-3}
			\end{tabu}$~~~~~~~~
   $\begin{tabu}{l|c||c|}
                \hline
                R_3 & ~~A~~ & P\\
                \hline
                \tau_5 & d & 0.9\\
                \tau_6 & e & 0.2\\
                \cline{2-3}
			\end{tabu}$}
\end{center}

         TID \ $(D^\prime)^p := \langle \mc{W}^{\prime},p^\prime \rangle$ is built as follows:    Introduce fresh constants $c_1,c_2,c_3$ for variables $x,y,z$ in $\mc{Q}$, and create  new endogenous tuples $\tau_1^U\!: \ R_1(c_1,c_2)$, $\tau_2^U\!: \ R_2(c_2)$, and $\tau_3^U\!: \ R_3(c_3)$; all with probability of $1$. The  \emph{fresh expansion} of $D$ via $\mc{Q}$ is $D^\prime = D \cup \{\tau_1^U, \tau_2^U,\tau_3^U\}$, with the distribution on $D$ extended to $D^\prime$ to make the latter a TID.  \
           Query $\mc{Q}$ has two components: $C_1=\{R_1(x,y) , R_2(y)\}$ and $C_2=\{R_3(z)\}$. \  With the   selection function \ $\sigma(C_1) := \tau_2^U$ and $\sigma(C_2) := \tau_3^U$, \ it holds: \  $P(\mc{Q},D^p) = \prod_{C \in \mc{C}(\mc{Q})} (1 - \nit{CE}(D^\prime,\mc{Q},\sigma(C)))
                        = (1 - \nit{CE}(D^\prime,\mc{Q},\tau_2^U)) \times (1 - \nit{CE}(D^\prime,\mc{Q},\tau_3^U)) 
                        =  (1 - 0.57) \times (1 - 0.02) = 0.3965$.
          \   With the  selection function \  $\sigma^\prime(C_1) := \tau_1^U$ and $\sigma^\prime(C_2) := \tau_3^U$, the result is the same since $\nit{CE}^I\!(D^\prime,\mc{Q},\tau_1^U) = \nit{CE}^I\!(D^\prime,\mc{Q},\tau_2^U)= 0.57$.
        \boxtheorem
    \end{example}

\vspace{-7mm}
       Notice that the extra tuples in Proposition  \ref{prop:connection_qep_gces}, although endogenous, behave as exogenous tuples in that they have probability $1$  (see Remark \ref{rem:exo}). 

\vspace{-2mm}

    \begin{theorem}\label{theo:complCQ} \em
Let $\mc{Q}$ be a BCQ without self-joins. \ (a) If
 $\mc{Q}$ is hierarchical, then, for every TID $D^p=\langle \mc{W}, p \rangle$ and endogenous tuple $\tau$, computing $\nit{CE}^{I\!}(D,\mc{Q},\tau)$ is in $\nit{PTIME}$.  \  (b) If $\mc{Q}$ is non-hierarchical, then computing  $\nit{CE}^I\!(D,\mc{Q},\tau)$ on TIDs is $\#P$-hard.\footnote{Meaning that computing  $\nit{CE}^I\!(D,\mc{Q},\tau)$ over inputs of the form $\langle D^p,\tau\rangle$, with $D^p$ a TID for $D$ and $\tau \in D^\nit{en}$, \ is a $\#P$-hard problem \citep{suciuGems}.}  \boxtheorem
    \end{theorem}

\vspace{-10mm}
 \begin{remark}\label{rem:afterP} \em \ (a) \ Theorem \ref{theo:complCQ} applies in particular to the computation of the CES on a TID with an arbitrary uniform distribution on endogenous tuples (and probability $1$ for exogenous tuples), and also to \ces, i.e. when endogenous tuples have probability $\frac{1}{2}$.\footnote{For discussions and results on the complexity of probabilistic query answering  for these different cases, see 
 \citep{amarilli,suciuGems,kenig21}).} 
 \ (b) \ Since \ces \  coincides with the BPI on DBs, we reobtain here, in a different way, the dichotomy result for BPI and BCQs \citep{Shapley_Tuple_Bertossi}. \
 (c) \ With some caveats, the results in this Section can be extended to UBCQs, but considering the notions of {\em safe} and {\em unsafe} queries \citep{dichotomy_UCQ} (see also \citep{suciuGems,benny22,sigRec23}). \ (c) For the hardness part of Theorem \ref{theo:complCQ} we used the existence of pseudo-exogenous tuples. \ignore{Using a  technique in \citep{benny22} in relation to elimination of exogenous tuples from the complexity proof of the Shapley-Value in DBs,} A natural question is whether it is possible to obtain the same result without appealing to them, but only to ``properly" endogenous tuples.
 \boxtheorem \end{remark}

\vspace{-7mm} As a consequence of the analysis in this section, it is clear that computing the Generalized Causal-Effect Score is also intractable.

\vspace{-2mm}
\section{Characterizing Properties of the GCES}
    \label{sec:ces_props}

    In this section we provide an axiomatic characterization of the Generalized Causal-Effect Score (GCES), that is, we show that this score is the only function satisfying a given set of properties. \ We consider the most general version of GCES, i.e.  for an arbitrary probability distribution over all possible worlds of an instance. However, we restrict ourselves to {\em monotone Boolean queries} (MBQs), which includes BCQ and  UBCQs.  The axioms depend on dealing with MBQs.

    In \citep{BPI_Math_Props}, an axiomatic characterization was given for the  Banzhaf Power Index (BPI). It is a general axiomatization in the context of game-theory. \ 
    As already mentioned, in \citep{Shapley_Tuple_Bertossi} it was shown that, in the context of query answering in DBs, \ces, i.e. the CES with independent tuples and  uniform distribution with parameter $\frac{1}{2}$ coincides with BPI; \ the latter applied with the endogenous tuples as players in a coalition game. \ From this connection, it follows that   there is an axiomatic characterization for this particular case of the CES.
    
In the following, we will show that only some of the characterizing properties, or axioms, for the BPI still hold for the GCES. We will also propose extra properties for the GCES, and will establish, in Theorem \ref{theo:gce_axioms}, that the  resulting set of axioms turns out to be {\em categorical}, i.e. fully characterizing for the GCES.\footnote{In Mathematical Logic, a {\em categorical theory}  has a single model (modulo isomorphism).} \ We will start introducing some useful  notions and notation that will allow us to formulate the characterizing properties of the GCES as appropriate for query answering in PDBs.

\vspace{-2mm}
    \begin{definition} \em \label{def:power}
      Let $\mc{Q}$ be a monotone Boolean query,  $D^p = \langle \mc{W}, p \rangle$ a PDB, and $W \in \mc{W}$. 

      \vspace{-3mm}
        \begin{itemize}
        \item[(a)] For \ $W \subsetneqq D^\nit{en}$ \  its \ \emph{power} is: 
        
        \vspace{-7mm}\begin{eqnarray*} 
        \nit{Power}(D^p,\mc{Q},W) &:=& \sum_{\tau \in D^\nit{en}} \Delta(\mc{Q},W,\tau), \ \ \mbox{with}\\
        \Delta(\mc{Q},W,\tau) &:=& \ \mc{Q}[W \cup D^\nit{ex} \cup \{\tau\}] - \mc{Q}[W \cup D^\nit{ex}].
        \end{eqnarray*}
        
 \vspace{-3mm}
        \item[(b)] For a tuple $\tau \in D^\nit{en}$, its \ {\em power} \ is: $$\nit{Power}(D^p,\mc{Q},\tau) \ := \sum_{W \  \subseteq \ (D^\nit{en} \smallsetminus \{\tau\})} \Delta(\mc{Q},W,\tau),$$ with $\Delta(\mc{Q},W,\tau)$ as in (a). 
 \vspace{-1mm}
    \item[(c)] The \emph{weighted power} of $\tau \in D^\nit{en}$ is:$$\nit{Power}^{w\!}(D^p,\mc{Q},\tau) \ := \sum_{W \subseteq (D^\nit{en} \smallsetminus \{\tau\})} \Delta(\mc{Q},W,\tau) \times p(W \cup D^\nit{ex}).$$
        
 \vspace{-1mm}
     \item[(d)] The \emph{total power} of the pair $(D^p,\mc{Q})$,  is:
$$\nit{Power}(D,\mc{Q}) \ := \sum_{W \ \subsetneqq \ D^\nit{en}} \nit{Power}(D^p,\mc{Q},W).$$ 
\end{itemize} \vspace{-1.2cm}\boxtheorem
    \end{definition}

\vspace{-3mm}  The \emph{power} of a tuple $\tau$ represents the number of subsets $W$ for which, adding $\tau$ to them, produces a change in the value of $\mc{Q}$. \ Similarly, the power of a subset $W$ represents  the number of tuples that, when added to $W$, change the query answer. \ 
  Notice that, equivalently, \ $\nit{Power}(D^p,\mc{Q}) = \sum_{\tau \in D^\nit{en}} \nit{Power}(D^p,\mc{Q},\tau)$.

    \begin{example} \label{ex:power} \em
        Consider the instance (with tuple-identifiers) $D^p=\{\tau_1\!\!:\!R(a,b), \ \tau_2\!\!:\!R(b,c), \ \tau_3\!\!:\!S(a,a), \ \tau_4\!\!:\!S(a,d)\}$ with exogenous tuples in $D^\nit{ex}\! =\! \{\tau_1\}$. For the BCQ \ $\mc{Q} \!: \exists x \exists y \exists z (R(x,y) \land S(x,z))$ \ and
        $W_1\! =\! \{\tau_2\}$: \  
            $\nit{Power}(D^p,\mc{Q},W_1) = \sum_{\tau \in \{\tau_2,\tau_3,\tau_4\}} \Delta(\mc{Q},W_1,\tau)
            = 0 + 1 + 1 = 2$.
     
        The \emph{power} of $\tau_2$ (as a tuple, not as the set $\{\tau_2\}$) is: \ 
        $\nit{Power}(D^p,\mc{Q},\tau_2) = \sum_{W \subseteq (D^\nit{en} \smallsetminus \{\tau_2\})}$ $ \Delta(\mc{Q},W,\tau_2) =\Delta(\mc{Q},\emptyset,\tau_1) + \Delta(\mc{Q},\{\tau_3\},\tau_2) + \Delta(\mc{Q},\{\tau_4\},\tau_2) + \Delta(\mc{Q},\{\tau_3,\tau_4\},\tau_2) = 0 + 0 + 0 + 0 = 0$. \   
        
        With $W_2\!=\! \{\tau_3\}, W_3\! =\! \{\tau_4\}, W_4\! =\! \{\tau_2,\tau_3\}, W_5\! =\! \{\tau_2,\tau_4\}, W_6\! =\! \{\tau_3,\tau_4\} , W_7\! =\! \{\tau_2,\tau_3,\tau_4\},$ $W_8 = \emptyset$, the \emph{power} of $W_i$ is $\nit{Power}(D^p,\mc{Q},W_i) = 2$ if $i\!=\!1,8$, but $0$, otherwise. Then, the \emph{total power} of $(D^p,\mc{Q})$ is: \ $\nit{Power}(D^p,\mc{Q}) = \sum_{W \subseteq D^\nit{en}} \nit{Power}(D^p,\mc{Q},W) = \nit{Power}(D^p,\mc{Q},W_1) \ + \cdots + \nit{Power}(D^p,\mc{Q},W_8) = 2 + 0 + \cdots + 0 + 2 = 4$. 
        
       As mentioned above,  the \emph{total power} can also be obtained through the \emph{power} of the tuples. \ The \emph{power} of $\tau_3$ and  $\tau_4$ are $\nit{Power}(D^p,\mc{Q},\tau_3) = \nit{Power}(D^p,\mc{Q},\tau_4) = 2$.  Then,  $\nit{Power}(D^p,\mc{Q}) = \sum_{\tau \in \{\tau_2,\tau_3,\tau_4\}} \nit{Power}(D^p,\mc{Q},\tau)  = 0 + 2 + 2 = 4$.
        \boxtheorem
    \end{example}
  
   \vspace{-4mm}{\em In the following, we will assume that we have a fixed PDB  $D^p = D^\nit{en} \cup D^\nit{ex}$, with $D^\nit{en} = \{\tau_1, \ldots, \tau_N\}$, and a  probability distribution $p$ on $\mc{W}$, its set of possible worlds. \ We restrict ourselves to the class $\msf{MBQ}$  of monotone Boolean queries for the schema of $D^p$.}

  Now, depending on a query $\mc{Q}$, each tuple in $D^\nit{en}$ will take a real number as a score that  represent its relevance for the answer to $\mc{Q}$ from $D^p$. \ This is represented as a  
 ``vectorial score-function",
$\psi$, from $\msf{MBQ}$ to $\mbb{R}^N$: \hspace{3mm} 
$\psi(\mc{Q}) \ = \ \langle \psi_{\tau_1}(\mc{Q}), \ldots, \psi_{\tau_N}(\mc{Q})\rangle$. 

\vspace{2mm}
Based on \citep{BPI_Math_Props},  we first analyze  the following potential properties of $\psi$:
    \begin{itemize}
    \item[]
    \begin{itemize}
            \item[{\sf DUM}:] \ (for ``dummy") \  If $\tau \in D^\nit{en}$ is a {\em dummy tuple}, meaning that  $\nit{Power}(D^p,\mc{Q},\tau) = 0$, then \ $\psi_{\tau}(\mc{Q}) = 0$.\footnote{For monotone queries, this condition is equivalent to  $\mc{Q}[W \cup D^\nit{ex}] =  \mc{Q}[W \cup D^\nit{ex} \cup \{\tau\}]$ for all $W \subseteq D^\nit{en} \smallsetminus \{\tau\}$, a more common formulation.}
            
            \item[{\sf EFF}:] \ (for ``efficiency") \  $\sum_{\tau \in D^\nit{en}} \psi_{\tau}(\mc{Q}) = \nit{Power}(D^p,\mc{Q})/(2^{N - 1})$.
            \item[{\sf SYM}:] \ (for ``symmetry") \  If $\mc{Q}[W \cup D^\nit{ex} \cup \{\tau\}] = \mc{Q}[W \cup D^\nit{ex} \cup \{\tau^\prime\}]$ for every $W \subseteq D^\nit{en} \smallsetminus \{\tau, \tau^\prime\}$, \ then \ $\psi_{\tau}(\mc{Q}) = \psi_{\tau^\prime}(\mc{Q})$.
            \item[{\sf LIN}:] \ (for ``linearity") \  For  MBQs $\mc{Q}$ and $\mc{Q}^\prime$, \ $\psi(\mc{Q} \lor \mc{Q}^\prime) + \psi(\mc{Q} \land \mc{Q}^\prime) = \psi(\mc{Q}) + \psi(\mc{Q}^\prime)$. \ignore{, where $\mc{Q} \land \mc{Q}^\prime$ and $\mc{Q} \lor \mc{Q}^\prime$ denote the conjunction and disjunction of the Boolean monotone queries, respectively.}
    \end{itemize}
    \end{itemize}

According to \citep{BPI_Math_Props}, as applied to our setting,  there is a unique function $\psi\!: \msf{MBQ} \rightarrow \mbb{R}^N$ that satisfies the properties {\sf DUM}, {\sf EFF}, {\sf SYM} and {\sf LIN}, and it 
corresponds to the BPI as defined in (\ref{eq:banzhaf}). For the formulation of the properties above, we do not use the possibly arbitrary distribution of the PDB $D^p$. However, the BPI on DBs coincides, in its turn, with the \ces \ case of the Causal-Effect Score in DBs \citep{Shapley_Tuple_Bertossi}. \ This leaves open the question about  properties that characterize GCES, the general case of the score, that we address in the rest of this section. We would expect the distribution $p$ of $D^p$ to play a role in them. 

It is worth mentioning that the Shapley value also satisfies {\sf DUM},  {\sf SYM} and {\sf LIN}, but not {\sf EFF}. However, it does satisfy slightly modified version of {\sf EFF}, where the sum of the value for all tuples is equal to $\mc{Q}[D]$ \ \citep{shapley1953original,aumann2015values}.

Example \ref{ex:power2} below shows that the GCES does not (always) satisfy {\sf SYM} and {\sf EFF}. It also illustrates that it  satisfies {\sf DUM} and {\sf LIN}. Actually, Theorem \ref{theo:gce_axioms} will show that the GCES always satisfies {\sf DUM}, {\sf LIN}, and modified versions of {\sf SYM} and {\sf EFF}.
   
    \ignore{For more intuition, consider any PDBs instance $\langle \mc{W}(D),p^D \rangle$ with $D$ an instance and $\tau \in D$ an endogenous tuple. If $\Delta(\mc{Q},W,\tau) = 0$ for all $W \subseteq D$ (dummy tuple), then $CE^{p^{\!D}}(D,\mc{Q},\tau) = 0$, and thus, GCES satisfies {\sf DUM}. However, we can easily show that {\sf EFF}, for an arbitrary PDB instance, is not satisfied: simply use a probability distribution $p^D$ different than the independent and uniform of $\frac{1}{2}$. The next example illustrates the latter.}

    \begin{example} \ \em (ex. \ref{ex:power} cont.)  \label{ex:power2} \
        Consider the PDBs $D^p = \langle \mc{W},p \rangle$ and $D^{p^\prime} = \langle \mc{W},p^\prime \rangle$ associated to the same instance  of Example \ref{ex:power}. \ Here, $N=3$, the number of endogenous tuples. 

        The probability distributions are defined as follows: \ $
        p(W_i \cup \{\tau_1\}) = 1/8$ for $i=1,\ldots,8$; \ and \ 
        $p^\prime(W_j \cup \{\tau_1\}) = 1/12$ \ for $j = 1,2,3,4$, \ and $p^\prime(W_k \cup \{\tau_1\}) = 1/6$ for $k = 5,6,7,8$; \ 
        and $p(W_i) = p^\prime(W_i) = 0$,  for $i = 1,\ldots,8$, \ because those $W_i$ do not contain $D^\nit{ex}$.
        
        Notice that $p$ creates a TID with distribution $U(\frac{1}{2})$, because: \  (a) for each $\tau \in D^\nit{en}$, $P(\tau) = \sum_{W, \tau \in W}p(W) = \frac{1}{2}$; \  and \ (b) for $W \subseteq D^\nit{en}$,   $p(W \cup D^\nit{ex}) =  \prod_{ \tau \in W} P(\tau) \times \prod_{ \tau \in (D^\nit{en} \smallsetminus W)} (1 - P(\tau))$ \ (see Section \ref{sec:backgrounda}). \ However, \ $p^\prime$ is neither independent nor uniform.

      First, we check {\sf DUM}. Notice that $\tau_2$ is the only \emph{dummy tuple}, i.e. $\nit{Power}(D^p,\mc{Q},\tau_2) = 0$, as in Example \ref{ex:power}. \ It holds $\nit{CE}(D^p,\mc{Q},\tau_2) = \nit{CE}(D^{p^\prime},\mc{Q},\tau_2) = 0$, and therefore, both distributions satisfy this property.
      
      Next, we check {\sf EFF} by computing the GCES of $\tau_3$ and $\tau_4$, for both PDBs, obtaining that $p$ satisfies {\sf EFF}, but  $p^\prime$ does not: \vspace{-2mm}
        \begin{align*}
            \sum_{\tau \in \{\tau_2,\tau_3,\tau_4\}} \nit{CE}(D^p,\mc{Q},\tau) \ &= \ 0 \ + \ 1/2 \ + \ 1/2 \ = \ \dfrac{\nit{Power}(D^p,\mc{Q})}{2^{N - 1}},\\
            \vspace{-4mm}
            \sum_{\tau \in \{\tau_2,\tau_3,\tau_4\}} \nit{CE}(D^{p^\prime},\mc{Q},\tau) \ &= \ 0 \ + \ 5/12 \ + \ 1/2 \ \neq \ \dfrac{\nit{Power}(D^{p^\prime},\mc{Q})}{2^{N - 1}}.
        \end{align*}
       
        Let us check {\sf SYM}. \ Tuples $\tau_3$ and $\tau_4$ should have the same GCES, because,  for every $W \subseteq \{\tau_2\}$, $\mc{Q}[W \cup D^\nit{ex} \cup \{\tau_3\}] = \mc{Q}[W \cup D^\nit{ex} \cup \{\tau_4\}]$. This holds for $p$, but not for $p^\prime$. 

\ignore{
        \comlb{Obvious?}
        \comfa{Es un poco obvio teniendo en cuenta que (1) $\tau_2$ es \textit{dummy tuple} y que (2) el CES para $\tau_3$ y $\tau_4$ está calculado implicitamente en las sumatorias de arriba (segundo y tercer sumando de cada sumatoria).}}
        
        Lastly, we check {\sf LIN}. \ Consider the query $\mc{Q}^\prime \!:\! \exists w S(w,w)$. We  only show that GCES satisfies {\sf LIN} for $\tau_3$ (it is similar for the other tuples). Computing its GCES with queries $Q^\prime$, $Q \wedge Q^\prime$ and $Q \lor Q^\prime$, we obtain: \   $\nit{CE}(D^p,\mc{Q}^\prime,\tau_3) = \nit{CE}(D^p,Q \wedge Q^\prime,\tau_3) = 1$, $\nit{CE}(D^p,Q \lor Q^\prime,\tau_3) = 1/2$; \ and $\nit{CE}(D^{p^\prime},\mc{Q}^\prime,\tau_3) = \nit{CE}(D^{p^\prime},\mc{Q}^\prime,\tau_3) = 1$, $ \nit{CE}(D^{p^\prime},\mc{Q}^\prime,\tau_3) = 5/12$. 
          \ignore{\begin{align*}
        \nit{CE}^{p^D}\!(D,\mc{Q}^\prime,\tau_3) = 1 ~~~&~~~ \nit{CE}^{(p^\prime)^D}\!(D,\mc{Q}^\prime,\tau_3) = 1,\\
        \nit{CE}^{p^D}\!(D,Q \wedge Q^\prime,\tau_3) = 1~~~&~~~ \nit{CE}^{(p^\prime)^D}\!(D,\mc{Q}^\prime,\tau_3) = 1,\\  
        \vspace{-4mm}
        \nit{CE}^{p^D}\!(D,Q \lor Q^\prime,\tau_3) = 1/2~~~&~~~ \nit{CE}^{(p^\prime)^D}\!(D,\mc{Q}^\prime,\tau_3) = 5/12,
    \end{align*}}
    It is easy to verify that {\sf LIN} is satisfied with both distributions.  \boxtheorem
    \end{example}
 
   \vspace{-7mm}This example motivates modifying  \ {\sf EFF} and {\sf SYM}, proposing their generalized versions:

\vspace{-2mm}
\begin{itemize}
    \item []
    \begin{itemize}
            \item[{\sf G-EFF}:] \ (for ``generalized-efficiency")  \
            \[ \sum_{\tau \in D^\nit{en}} \!\! \psi_{\tau}(\mc{Q}) =  \!\!\sum_{W \subsetneqq D^\nit{en}} \sum_{\tau \in (D^\nit{en} \smallsetminus W)} \!\!\!\! \Delta(\mc{Q},W,\tau) \times \left( p(W\cup D^\nit{ex}) + p(W\cup D^\nit{ex} \cup \{\tau\}) \right).\footnote{The sums on the RHS can be given in terms of $W \subseteq D^\nit{en}$ and $\tau \in D^\nit{en}$, because $\Delta(Q,W,\tau) = 0$ for every $\tau \in W$.}\]
            \item[{\sf G-SYM}:] \ (for ``generalized-symmetry'') \
            If $\Delta(\mc{Q},W,\tau) = \Delta(\mc{Q},W,\tau^\prime) = 0$ for every $W \subseteq D^\nit{en} \smallsetminus \{\tau, \tau^\prime\}$, \ then \ $\psi_{\tau}(\mc{Q}) - \nit{Power}^{w\!}(D^p,\mc{Q},\tau) \ = \ \psi_{\tau^\prime}(\mc{Q}) - \nit{Power}^{w\!}(D^p,\mc{Q},\tau^\prime).$\footnote{One could argue that the property should also hold under the ``symmetric" condition ``$\Delta(\mc{Q},W,\tau) = \Delta(\mc{Q},W,\tau^\prime) = 1$ for all possible $W \subseteq D^\nit{en} \smallsetminus \{\tau, \tau^\prime\}$". However, if that were the case, we would have $Q[W] = 0$ and $Q[W \cup \{\tau\}] = Q[W \cup \{\tau^\prime\}] = 1$. Since $W \cup \{\tau\}$ and $W \cup \{\tau^\prime\}$ are distinct sets, possibly with different probabilities, it is not sensible to impose this kind of symmetry. \ignore{A change in the formulation of this condition will imply a change on the property and the proof of the Theorem.}}
    \end{itemize}
\end{itemize}

{\sf G-EFF} is similar to {\sf EFF}, but involves an arbitrary distribution $p$. If the latter is independent and uniform $U(\frac{1}{2})$, then $\left( p(W\cup D^\nit{ex}) + p(W\cup D^\nit{ex} \cup \{\tau\}) \right)$ becomes $1/2^{N-1}$ ($N = |D^\nit{en}|$), for every $\tau \in D^\nit{en}$ and $W \subsetneqq D^\nit{en}$. 

The equalities in the consequents of {\sf G-SYM} and {\sf SYM} are similar, but, for the former, we remove the tuples' powers. With an independent and $U(\frac{1}{2})$ distribution, if two tuples satisfy the hypothesis, the term $\nit{Power}(D^p,\mc{Q},\tau)$ is the same for both, reobtaining {\sf SYM}.

    \begin{theorem} \em
        \label{theo:gce_axioms}
        Let \ $D^p = \langle \mc{W},p\rangle$ \ be a PDB. \
        There is a unique score function $\psi$ from $\msf{MBQ}$, the class of monotone Boolean queries,  to $\mbb{R}^{N}$, with $N = |D^\nit{en}|$, that satisfies the properties {\sf DUM}, \ {\sf G-EFF}, \ {\sf G-SYM} \ and \ {\sf LIN}. \ Moreover, this function coincides with the GCES. \boxtheorem
    \end{theorem}

       \vspace{-7mm}
    This theorem does not make any assumption on the distribution $p$. In particular, neither tuple-independence nor uniformity are required.
\   It is worth noticing that the property of monotonicity of the queries is used in the proof of the Theorem, particularly for uniqueness, and for GCES' satisfaction of  \textsf{G-SYM}. 
   \ignore{We do not discard a possible axiomatic characterization of GCES for non-monotone queries, however, the set of properties to be considered may differ from the one here presented.}

\vspace{-2mm}
    \section{Discussion and Conclusions}\label{sec:conclusions}

\vspace{-1mm}
     In between, and in parallel work, we have managed to investigate in detail the conditions under which \ces \ (i.e. GCES with independence and $U(\frac{1}{2})$ distribution), Shapley and RESP are aligned, i.e. they produce the same rankings (a property also called ``ordinal equivalence" \citep{freixas}). \ For some of our results see \citep{sigmod24}. \ Experimentation with cases of queries whose scores are not aligned is left for ongoing and future work, so as other open research directions we mention next. 
     
{\em Beyond Monotone Queries:} \ We have restricted ourselves to monotone queries, and this assumption play an important role in the characterization of the GCES. It is left open to find a characterization for interesting classes of non-monotone queries, e.g. conjunctive with negative literals.

{\em Beyond Boolean Queries:} \ For the axiomatic characterization of the GCES, we have restricted ourselves to Boolean queries. It would be interesting to have such an axiomatization for more general monotone numerical queries, e.g. some aggregate queries.

{\em  GES Monotonicity:} \ It would be interesting to investigate more deeply the monotonicity properties (or the lack thereof) of the GCES. \ For example, under what changes in the database, or in the tuple probabilities, the GCES of  tuples change accordingly (or the other way around).
    
    {\em Score Computation:} \ Due to the intrinsic high complexity of computing the  GCES, it is worth exploring efficient approximate algorithms, possibly for some interesting cases of queries. 

 {\em Aggregate Queries:} \ In this work, we considered mostly conjunctive queries. A natural extension is a deeper investigation of queries with aggregations on CQs. We made early on the case for the convenience of the GCES for this kind of queries.   
    
{\em Semantic Constraints:} \ RESP has been formalized and investigated in the presence of integrity constraints, and its behavior changes \citep{flairs17}. This is something to investigate for the GCES, in its basic and generalized versions. Particularly interesting becomes dealing with constraints on probabilistic DBs \citep{suciuGems}. 

{\em  GES Robustness:} \ It would be interesting to analyze the {\em robustness} of the Generalized CES, under small variations of  parameters, such as the distribution of the probabilistic DB.

{\em Attribute-Level GCES:} \ We defined the GCES at the tuple level. It would be interesting to extend its definition and investigation in order to quantify the causal effect of an attribute value in a tuple. \ This case is challenging in that it is not only about making an attribute value true or false anymore. Making the latter false may lead to consider multiple alternative values, as has been done for RESP \citep{deem,adbis23}. 

\ignore{{\em CES for Explainable ML:} \ So as the Shapley value (as SHAP) and RESP have been used to provide attributive explanations in Machine Learning (see \citep{adbis23} for references), it would be interesting to explore the applicability of the GCES in ML, for the same purpose.}

\acks{Felipe Azua has been supported by the Millennium Institute for Foundational Research on Data (IMFD, Chile). Part of this work was done while Felipe Azua was visiting  LIMOS (U. Clermont-Ferrand, France). \
 Leopoldo Bertossi has been supported by NSERC-
DG 2023-04650, and the  IMFD. \ Useful comments by one of the anonymous reviewers are extremely appreciated.}


\newpage
\appendix

\section*{Appendix: \ Proofs}

\section{Proofs of Section 4}

   \noindent {\bf Proof of Proposition \ref{prop:connection_qep_gces}:} \
    Let $D^p = \langle \mc{W}, p \rangle$ be a TID associated to relational instance with $D$. Now, consider a \emph{fresh expansion} of $D$ over $\mc{Q}$, $D^\prime = D \cup \mc{T}$. Each new tuple $\tau^U \in \mc{T}$ will have a probability of 1 and will be independent from every other tuple. The tuples in $D \cap D^\prime$ retain there original (independent) probabilities. \
    Now, consider a new tuple $\tau^U \in \mc{T}$ created from an atom of the component $C_k \in \mc{C}(\mc{Q})$, we can compute its CES using the new probability distribution $p^\prime$: \ $
    \nit{CE}^{I\!}(D^\prime,\mc{Q},\tau^U) 
        = \mbb{E}(\mc{Q}~|~\nit{do}(\tau^U \In)) - \mbb{E}(\mc{Q}~|~\nit{do}(\tau^U \Out))
        = P(\mc{Q}~|~\nit{do}(\tau^U \In)) - P(\mc{Q}~|~\nit{do}(\tau^U \Out))
        = 1 - P(\mc{Q}~|~\nit{do}(\tau^U \Out))$, \ the last probability corresponds to interventions on $p^\prime$. \
        
    It holds: (a) $P(\mc{Q}, D^\prime) = 1$ since $(D^\prime \smallsetminus D) \models \mc{Q}$, that is, the set of new tuples $\tau^U$ is sufficient to make the query true, and each new tuple has a probability of 1.
    \ This probability will not change by the intervention $do(\tau^U \In)$. \ (b) By Remark \ref{rem:ind_probability}, we can compute $P(\mc{Q}~|~\nit{do}(\tau^U \Out))$ by multiplying the probability of the sub-query of each component of $\mc{Q}$, and since each component, will have a probability of 1 (because each tuple in $\mc{T}$ has a probability of 1), $P(\mc{Q}~|~\nit{do}(\tau^U \Out)) = P(\mc{Q}_k,D)$, with $\mc{Q}_k$ the sub-query of component $C_k$. Note that, if $\tau^U, \tau^{U^\prime} \in \mc{T}$ were created from atoms of the same component, say $C_k$, then $\nit{CE}^{I\!}(D^\prime,\mc{Q},\tau^U) = \nit{CE}^{I\!}(D^\prime,\mc{Q},\tau^{U^\prime})$. \
    Consider a selection function $\sigma$ as in Definition \ref{def:component}. Then, by Remark \ref{rem:ind_probability}, we can rewrite the probability of the subquery of each component as $P(\mc{Q}_i,D) = 1 - \nit{CE}^{I}(D,\mc{Q},\sigma(C_i))$. It follows that \
    $P(\mc{Q},D^p) \ = \ \prod_{C \in \mc{C}(\mc{Q})} (1 - \nit{CE}^{I}(D,\mc{Q},\sigma(C)))$.
    Since two tuples from atoms of the same component will have the CES, then any selection function $\sigma$ would produce the same result.
    \boxtheorem

  \noindent {\bf Proof of Theorem \ref{theo:complCQ}:}  \  Assume $\mc{Q}$ is hierarchical. Its causal effect is given in Remark \ref{rem:gcs}(a). Since $D^p$ is a TID, the two probabilities there are also for TIDs (see Remark \ref{rem:tid}(b)). As a consequence, they can be computed in polynomial-time.

Assume $\mc{Q}$ is non-hierarchical. We appeal to Proposition \ref{prop:connection_qep_gces}. Let $D^p$ be a TID that makes computing $P(\mc{Q},D^p)$  a $\#P$-hard case. \ Now, for $D^p$, consider the TID   $(D^\prime)^p$ as in Proposition \ref{prop:connection_qep_gces}, for which there are  endogenous tuples satisfying (\ref{eq:prod}), which shows that there are tuples for which computing $\nit{CE}^I$ is as hard as computing $P(\mc{Q},D^p)$.
 \boxtheorem

\section{Proof of Theorem \ref{theo:gce_axioms}}
 In order to prove Theorem \ref{theo:gce_axioms}, we need some notions and a  technical result. \ First, we need to introduce a particular query and its notation. For a fixed instance $D$, and  $S \subseteq D$,  \ $\mc{Q}_S$ denotes the following monotone Boolean query:\footnote{Notice that, since $S$ is fixed and finite, it can be expressed in the FO language of the schema.  It can be written as the conjunction of the tuples, as ground atoms, of the set $S$.} \ For $W \in \mc{W}$, 
    \begin{equation} \label{eq:query_qs}
        \mc{Q}_S[W] \ := \ 
            \begin{cases}
                1 & \text{ , if } \ S \subseteq W\\
                0 & \text{ , otherwise.}
            \end{cases}
    \end{equation}

    \begin{example} \ \em (ex. \ref{ex:power} cont.)  \
      Query $\mc{Q}_{W_3}$ is defined by  \ $ \mc{Q}_{W_3}[W] := 1 $ \ if \ $W_3 \subseteq W$; and $0$ otherwise. \ 
       It  can be expressed as a the conjunction of the tuples (atoms) in  $W_3$: \ \  $\mc{Q}_{W_3}\!\!: (R(a,b) \land S(b))$. \boxtheorem
    \end{example}

 \begin{definition} \em \label{def:mss}
Consider a MBQ $\mc{Q}$ and an instance $D$. A  $W \subseteq D$ is a \emph{minimal satisfiable set} if $\mc{Q}[W] = 1$, and $\mc{Q}[W \smallsetminus \{\tau\}] = 0$, for every $\tau \in W$. \ $\nit{MSS}(D,\mc{Q})$ denotes the set of all minimal satisfiability sets for $D$ and $\mc{Q}$. \boxtheorem
\end{definition}

    \begin{lemma}\em
        \label{lem:query_decom}
        Let $D$ be an instance, $\mc{Q}$ be a MBQ, and $\nit{MSS}(\mc{Q}) = \{S_1, \ldots, S_m\}$.  \ $\mc{Q}$ \ can be expressed as follows: \ For $W \in \mc{W}(D)$, \ $
        \mc{Q}[W] \ = \ (\mc{Q}_{S_1} \lor \dotsb \lor \mc{Q}_{S_m})[W]$, \ 
      with the queries $\mc{Q}_{S_i}$ defined as in  (\ref{eq:query_qs}). 
    \end{lemma}

\vspace{2mm}
\noindent {\bf Proof of Lemma \ref{lem:query_decom}:} \ 
Consider a MBQ $\mc{Q}$ and an instance $D$. The statement of the lemma is equivalent to: \   $\mc{Q}[W] = 1$ iff there exists $S_i \in \nit{MSS}(D,\mc{Q})$, such that $S_i \subseteq W$. \ Now we prove this claim (both directions) by contradiction.

        \vspace{1mm}
\noindent ($\Rightarrow$) \ Let $W^{\star}$ be a set such $\mc{Q}[W^{\star}] = 1$, but, for every $S_i \in \nit{MSS}(\mc{Q})$, \   $S_i \not\subseteq W^\star$. \  Two cases arise: (a) If $\mc{Q}[W^{\star} \smallsetminus \{\tau\}] = 0$ for every $\tau \in W^{\star}$, then $W^{\star} \in \nit{MSS}(D,\mc{Q})$; \ and \ (b) There is a tuple $\tau$, such that $\mc{Q}[W^{\star} \smallsetminus \{\tau\}] = 1$. \ Case (a) is clearly a contradiction. \ In case (b), we can remove tuples from $W^{\star}$ until case (a) occurs, eventually leading to a contradiction. \ Therefore, such a $W^{\star}$ cannot exist. 

\noindent ($\Leftarrow$) \ Assume that there exists $W \subseteq D$ such $\mc{Q}[W] = 0$, \ and $S \subseteq W$ with $S \in \nit{MSS}(D,\mc{Q})$. Since $S \in \nit{MSS}(D,\mc{Q})$, \ $\mc{Q}[S] = 1$, which cannot happen, because $\mc{Q}$ is monotone. So, such a  $W$ cannot exist.
\boxtheorem

 \ignore{ \vspace{-1mm} The lemma tells us that  all Boolean monotone queries $\mc{Q}$ can be rewritten as a disjunction of its minimal \red{swinging sets.}}

   \vspace{2mm}
    \noindent {\bf Proof of Theorem \ref{theo:gce_axioms}:\footnote{This proof  is inspired by a proof in \citep{BPI_Math_Props} for the BPI.}}  \
        First, we prove the uniqueness of the function $\psi$, and then, that the GCES satisfies all properties. \ 
  Consider \ignore{$D^p = \langle \mc{W}(D),p^D \rangle$ a PDB, with $D$ an (non-probabilistic) instance and} a monotone Boolean query $\mc{Q}$, and its  minimal satisfiable sets, say $\nit{MSS}(D,\mc{Q}) = \{S_1, \ldots, S_m\}$.  By Lemma \ref{lem:query_decom}, \ignore{any monotone Boolean query} $\mc{Q}$ can be decomposed in queries of the form  (\ref{eq:query_qs}). \ That is, \ for every $W \subseteq D$: \ $\mc{Q}[W] = (\mc{Q}_{S_1} \lor \cdots \lor \mc{Q}_{S_m})[W]$.

        Now, consider any of the individual queries in the disjunction, say \  $\mc{Q}_{S_i}$.  By property {\sf G-EFF}, the following holds:
        \[
        \sum_{\tau \in D^\nit{en}} \!\! \psi_{\tau}(\mc{Q}) =  \!\!\sum_{W \subsetneqq D^\nit{en}} \sum_{\tau \in (D^\nit{en} \smallsetminus W)} \!\!\! \Delta(\mc{Q},W,\tau) \times \left( p(W\cup D^\nit{ex}) + p(W\cup D^\nit{ex} \cup \{\tau\}) \right).
        \]
        Three observations: \ 
        (a) We can express $\sum_{\tau \in D^\nit{en}} \psi_{\tau}(\mc{Q}_{S_i})$ as:
        \begin{align*}
            \sum_{\tau \in D^\nit{en}} \psi_{\tau}(\mc{Q}_{S_i}) =& \sum_{\tau \in D^\nit{en}} \sum_{W \subseteq (D^\nit{en} \smallsetminus \{\tau\})} \Delta(\mc{Q}_{S_i},W,\tau) \times p(W \cup D^\nit{ex}) \ \ +  \\ 
            &\sum_{\tau \in D^\nit{en}} \sum_{W \subseteq (D^\nit{en} \smallsetminus \{\tau\})}  \Delta(\mc{Q}_{S_i},W,\tau) \times p(W \cup D^\nit{ex} \cup \{\tau\}). 
        \end{align*}
        (b) 
        \[
        \sum_{\tau \in D^\nit{en}} \nit{Power}^{w\!}(D^p,\mc{Q}_{S_i},\tau) = \sum_{\tau \in D^\nit{en}} \sum_{W \subseteq (D^\nit{en} \smallsetminus \{\tau\})} \Delta(\mc{Q}_{S_i},W,\tau) \times p(W \cup D^\nit{ex}).
        \]
        
        \vspace{1mm} \noindent (c) \ For any two tuples $\tau,\tau^\prime \in (S_i \cap D^\nit{en})$, \  $\Delta(\mc{Q}_{S_i},W,\tau) = \Delta(\mc{Q}_{S_i},W,\tau^\prime) = 0$, whenever $W \subseteq (D^\nit{en} \smallsetminus \{\tau,\tau^\prime\})$. Therefore, by {\sf G-SYM},\ it holds:
        \[
        \psi_{\tau}(\mc{Q}_{S_i}) - \nit{Power}^{w\!}(D^p,\mc{Q}_{S_i},\tau) \ = \ \psi_{\tau^\prime}(\mc{Q}_{S_i}) - \nit{Power}^{w\!}(D^p,\mc{Q}_{S_i},\tau^\prime).
        \]
       
        From (a) and (b), we obtain: \  $\sum_{\tau \in D^\nit{en}} (\psi_{\tau}(\mc{Q}_{S_i}) - \nit{Power}^{w\!}(D^p,\mc{Q}_{S_i},\tau)) = k$, where $k$ is
        \[
        k = \sum_{\tau \in D^\nit{en}} \sum_{W \subseteq (D^\nit{en} \smallsetminus \{\tau\})} \Delta(\mc{Q}_{S_i},W,\tau) \times p(W \cup D^\nit{ex} \cup \{\tau\})
        \].

Furthermore, each endogenous tuple $\tau \not\in S_i$ is \emph{dummy}, and, therefore, by property {\sf DUM}, $\psi_{\tau}(\mc{Q}_{S_i}) = 0$. \ Then, \ by (c): \  $\psi_{\tau}(\mc{Q}_{S_i}) - \nit{Power}^{w\!}(D^p,\mc{Q}_{S_i},\tau) = \frac{k}{|D^\nit{en} \cap S_i|}$. This expression uniquely defines the value of function $\psi_{\tau}$ for query $\mc{Q}_{S_i}$:
        \[
        \psi_{\tau}(\mc{Q}_{S_i}) = \begin{cases}
            \frac{k}{|D^\nit{en} \cap S_i|} + \nit{Power}^{w\!}(D^p,\mc{Q}_{S_i},\tau) & \text{ , if } \ \tau \in S \cap D^\nit{en} \\
            0 & \text{ , otherwise. }
        \end{cases}
        \]
        Now, by property {\sf LIN}, we can recursively obtain for $\psi(\mc{Q})$:
        \[
        \psi(\mc{Q}) = \psi(\mc{Q}_l \lor \mc{Q}_r) = \psi(\mc{Q}_l) + \psi(\mc{Q}_r) - \psi(\mc{Q}_l \land \mc{Q}_r),
        \]
        where $\mc{Q}_l[W] := (\mc{Q}_{S_1} \lor \cdots \lor \mc{Q}_{S_k})[W]$ and $\mc{Q}_r[W] := (\mc{Q}_{S_{k+1}} \lor \cdots \lor \mc{Q}_{S_m})[W]$, with $1\leq k < m$ and $W \subseteq D^p$.\footnote{Any query of the form $\mc{Q}_{S_i} \land \mc{Q}_{S_j}$ can be expressed as $\mc{Q}_{S_i \cap S_j}$.} Then, as each $\psi_{\tau}(\mc{Q}_{S_i})$ is uniquely defined for any $S_i \in \nit{MSS}(D^p,\mc{Q})$ and $\tau \in D^\nit{en}$, we conclude that $\psi_{\tau}(\mc{Q})$ is uniquely defined for each $\tau \in D^\nit{en}$.

     Having established uniqueness, we now turn to proving that the GCES satisfies all the given properties for a monotone query $\mc{Q}$ and a PDB $D^p$ with  distribution $p$. \ First, notice that, for a given MBQ $\mc{Q}$, and an endogenous tuple $\tau \in D^p$, the GCES for can be written as:
        \begin{equation}
            \label{GCE:v2}
            \nit{CE}(D^p,\mc{Q},\tau) \ = \sum_{W \subseteq (D^\nit{en}\smallsetminus\{\tau\})}  \Delta(\mc{Q},W,\tau) \times \left(p(W \cup D^\nit{ex}) + p(W \cup D^\nit{ex} \cup \{\tau\})\right).
        \end{equation}
        The property {\sf DUM} is trivial, since if a tuple $\tau$ does not contribute to any tuple, then $\mc{Q}[W] - \mc{Q}[W \smallsetminus \{\tau\}] = 0$ for every world $W$. 
        
        For property {\sf G-EFF}, we can group the sum of all the scores by the subsets of $D^p$, that is:
        \begin{align*}
            \sum_{\tau \in D^\nit{en}} \nit{CE}(D^p,\mc{Q},\tau) &= \sum_{\tau \in D^\nit{en}} \sum_{W \subseteq (D^\nit{en}\smallsetminus\{\tau\})}  \Delta(\mc{Q},W,\tau) \times \left(p(W \cup D^\nit{ex}) + p(W \cup D^\nit{ex} \cup \{\tau\})\right) \\
            &= \sum_{W \subsetneqq D^\nit{en}} \sum_{\tau \in D^\nit{en}}  \Delta(\mc{Q},W,\tau) \times \left(p(W \cup D^\nit{ex}) + p(W \cup D^\nit{ex} \cup \{\tau\})\right).
        \end{align*}  
        Since $\Delta(\mc{Q},W,\tau) = 0$ for every $\tau \in W$, we obtain the desired expression for {\sf G-EFF}.

        For property {\sf G-SYM}, consider that:
        \begin{align*}
            \nit{CE}(D^p,\mc{Q},\tau) - \nit{Power}^{w\!}(D^p,\mc{Q}_{S_i},\tau) = \sum_{W \subseteq (D^\nit{en} \smallsetminus \{\tau\})} \Delta(\mc{Q},W,\tau) \times p(W \cup D^\nit{ex} \cup \{\tau\})
        \end{align*}
        Then, to verify {\sf G-SYM}, we need to show that, for any two tuples $\tau, \tau^\prime \in D^\nit{en}$ satisfying $\Delta(\mc{Q},W,\tau) = \Delta(\mc{Q},W,\tau^\prime) = 0$ for every $W \subseteq D^\nit{en} \smallsetminus \{\tau,\tau^\prime\}$, the following holds:
        \begin{equation}
        \sum_{W \subseteq D^\nit{en} \smallsetminus \{\tau\}} \!\!\!\!\!\Delta(\mc{Q},W,\tau) \times p(W \cup D^\nit{ex} \cup \{\tau\}) \ = \!\!\!\sum_{W \subseteq (D^\nit{en} \smallsetminus \{\tau^\prime\})} \!\!\!\!\Delta(\mc{Q},W,\tau^\prime) \times p(W \cup D^\nit{ex} \cup \{\tau^\prime\}). \label{eq:fla}
        \end{equation}
        Notice that, for $\tau$, $\Delta(\mc{Q},W,\tau) = \Delta(\mc{Q},W,\tau^\prime) =  0$ for all $W \subseteq (D^\nit{en} \smallsetminus \{\tau,\tau^\prime\})$. \ By this, we can express the sums as:
        \begin{multline*}
            \sum_{W \subseteq (D^\nit{en} \smallsetminus \{\tau,\tau^\prime\})} \Delta(\mc{Q},W \cup \{\tau^\prime\},\tau) \times p(W \cup \{\tau^\prime\} \cup D^\nit{ex} \cup \{\tau\}) \ = \\  \sum_{W \subseteq (D^\nit{en} \smallsetminus \{\tau,\tau^\prime\})} \Delta(\mc{Q},W\cup \{\tau\},\tau^\prime) \times p(W \cup \{\tau\} \cup D^\nit{ex} \cup \{\tau^\prime\}).
        \end{multline*}
        From  the initial condition $\Delta(\mc{Q},W,\tau) = \Delta(\mc{Q},W,\tau^\prime)$ for every $W \subseteq D^\nit{en} \smallsetminus \{\tau,\tau^\prime\}$, and the fact that $\mc{Q}$ is monotone, we obtain $\mc{Q}[W \cup D^\nit{ex} \cup \{\tau\}] = \mc{Q}[W \cup D^\nit{ex} \cup \{\tau^\prime\}]$. From this, it follows $\Delta(\mc{Q},W \cup \{\tau^\prime\},\tau) = \Delta(\mc{Q},W \cup \{\tau\},\tau^\prime)$ for every $W \subseteq D^\nit{en}\smallsetminus \{\tau,\tau^\prime\}$. We obtain that the equality (\ref{eq:fla}) holds, and therefore, GCES satisfies {\sf G-SYM}.
        
        \ignore{Then, $\Delta(\mc{Q},W \cup \{\tau^\prime\},\tau) = \Delta(\mc{Q},W \cup \{\tau\},\tau^\prime)$ for all $W \subseteq D^\nit{en}\smallsetminus \{\tau,\tau^\prime\}$ since $\mc{Q}[W \cup D^\nit{ex} \cup \{\tau\}] = \mc{Q}[W \cup D^\nit{ex} \cup \{\tau^\prime\}]$\footnote{This is true because of: (a) the initial condition $\Delta(\mc{Q},W,\tau) = \Delta(\mc{Q},W,\tau^\prime)$ for any $W \subseteq D^\nit{en} \smallsetminus \{\tau,\tau^\prime\}$, and (b) $\mc{Q}$ is monotone.}. It follows that the equality holds and therefore, GCES satisfies {\sf G-SYM}.}
        
        For {\sf LIN}, let $\mc{Q}$ and $\mc{Q}^\prime$ be two monotone Boolean queries. Notice that the expression $p(W \cup D^\nit{ex}) + p(W \cup D^\nit{ex} \cup \{\tau\})$ does not depend on the query, only on the possible world $W$. Then, it is immediate that, for a given PDB instance $D^p$ and any given endogenous tuple $\tau \in D^\nit{en}$, the following holds:

       \vspace{1mm}
       $\nit{CE}(D^p,\mc{Q} \land \mc{Q}^\prime,\tau) + \nit{CE}(D^p,\mc{Q} \lor \mc{Q}^\prime,\tau) = \nit{CE}(D^p,\mc{Q},\tau) + \nit{CE}(D^p,\mc{Q}^\prime,\tau).$
     {\boxtheorem}

\end{document}